\documentclass[twocolumn]{aastex63}
\usepackage{amsmath}

\usepackage{mathrsfs}

\received{November 27, 2019}
\revised{March 27, 2020}
\accepted{March 30, 2020}
\submitjournal{ApJS}

\shorttitle{Estimating luminosity functions via KDE}
\shortauthors{Yuan et al.}

\begin{document}

\title{A flexible method for estimating luminosity functions via Kernel Density Estimation}

\correspondingauthor{Zunli Yuan}
\email{yuanzunli@ynao.ac.cn}

\author[0000-0001-6861-0022]{Zunli Yuan}
\affiliation{Yunnan Observatories, Chinese Academy of Sciences, Kunming 650216, P. R. China}
\affiliation{Key Laboratory for the Structure and Evolution of Celestial Objects, Chinese Academy of Sciences, Kunming 650216, P. R. China}
\affiliation{Center for Astronomical Mega-Science, Chinese Academy of Sciences, Beijing 100012, P. R. China}

\author[0000-0001-7039-9078]{Matt J. Jarvis}
\affiliation{Astrophysics, University of Oxford, Denys Wilkinson Building, Keble Road, Oxford, OX1 3RH, UK}
\affiliation{Physics and Astronomy Department, University of the Western Cape, Bellville, 7535, South Africa}

\author{Jiancheng Wang}
\affiliation{Yunnan Observatories, Chinese Academy of Sciences, Kunming 650216, P. R. China}
\affiliation{Key Laboratory for the Structure and Evolution of Celestial Objects, Chinese Academy of Sciences, Kunming 650216, P. R. China}
\affiliation{Center for Astronomical Mega-Science, Chinese Academy of Sciences, Beijing 100012, P. R. China}



\begin{abstract}

We propose a flexible method for estimating luminosity functions (LFs) based on kernel density estimation (KDE), the most popular
nonparametric density estimation approach developed in modern statistics, to overcome issues surrounding binning of LFs.
One challenge in applying KDE to LFs is how to treat the boundary bias problem, since astronomical surveys usually
obtain truncated samples predominantly due to the flux-density limits of surveys.
We use two solutions, the transformation KDE method ($\hat{\phi}_{\mathrm{t}}$), and the transformation-reflection KDE method ($\hat{\phi}_{\mathrm{tr}}$)
to reduce the boundary bias. We develop a new likelihood cross-validation criterion for selecting optimal bandwidths, based on which,
the posterior probability distribution of bandwidth and transformation parameters for $\hat{\phi}_{\mathrm{t}}$ and $\hat{\phi}_{\mathrm{tr}}$
are derived within a Markov chain Monte Carlo (MCMC) sampling procedure.
The simulation result shows that $\hat{\phi}_{\mathrm{t}}$ and $\hat{\phi}_{\mathrm{tr}}$ perform better than
the traditional binned method, especially in the sparse data regime around the flux-limit of a survey or at the bright-end of the LF.
To further improve the performance of our KDE methods, we develop the transformation-reflection adaptive KDE approach ($\hat{\phi}_{\mathrm{tra}}$).
Monte Carlo simulations suggest that it has a good stability and reliability in performance,
and is around an order of magnitude more accurate than using the binned method.
By applying our adaptive KDE method to a quasar sample, we find that it achieves estimates comparable to the rigorous determination
 in a previous work, while making far fewer assumptions about the LF. The KDE method we develop has the advantages of both parametric and
non-parametric methods.

\end{abstract}




\section{Introduction}
\label{Intro}
Redshift and luminosity (or magnitude) are undoubtedly the most two fundamental observational quantities for galaxies.
As a bivariate density function of redshift and luminosity, the luminosity function (LF) provides
one of the most important tools to probe the distribution and evolution of galaxies and AGNs (active galactic nuclei) over cosmic
time. Measuring the LF at various wavebands has long been an important pursuit
of a large body of work over the last 50 years \citep[e.g.,][]{1976ApJ...203..297S,1988MNRAS.232..431E,1990MNRAS.247...19D,2000MNRAS.317.1014B,
2001MNRAS.322..536W,2001MNRAS.326..255C,2001MNRAS.327..907J,2003ApJ...598..886U,2003MNRAS.339.1057Y,2006AJ....131.2766R,2007ApJ...665..265F,2010MNRAS.401.2531A,
2012ApJ...751..108A,2015ApJ...803...34B,2015MNRAS.452.1817B,2016ApJ...829...33Y,2016ApJ...820...65Y,2019MNRAS.488.1035K,2020arXiv200107727T}.
However, accurately constructing the LF remains a challenge, since the observational selection
effects (e.g. due to detection thresholds in flux density, apparent magnitude, colour, surface
brightness, etc.) lead truncated samples and thus introduce bias into the LF estimation \citep{2011A&ARv..19...41J}.

Up to now there have been numerous statistical approaches devised to measure the LF.
These mainly include parametric techniques and non-parametric methods \citep[see][for an overview]{2011A&ARv..19...41J}.
Parametric methods typically provide an analytic form for the estimated LF,
where the parameters are determined by maximum likelihood estimator
\citep[e.g.,][]{1979ApJ...232..352S,1983ApJ...269...35M,2005MNRAS.360..727A,2000MNRAS.319..121J,2001MNRAS.322..536W,2019ICRC...36..638B},
or within a Bayesian framework and applying a Markov Chain Monte Carlo (MCMC)
algorithm \citep[e.g.,][]{2006MNRAS.369..969A,2014MNRAS.441.1760Z,2016ApJ...820...65Y,2017ApJ...846...78Y}.
The parametric method can naturally incorporate various complicated selection
functions, and has the flexibility to be extended to trivariate LFs
which need to define additional quantities, such as photon index
in the case of $\gamma$- and X-ray LFs \citep[e.g.][]{2012ApJ...751..108A} and
spectral index for radio LFs \citep[e.g.][]{2000MNRAS.319..121J,2016ApJ...829...95Y},
aside from luminosity and redshift. In addition, with the help of powerful statistical techniques such as the MCMC algorithm, the parametric method can be very accurate.
But the premise is that we are fortunately enough to ``guess'' the right functional form for the LF. To reproduce the features of ``observed LFs''
\footnote{The so-called "observed LFs" are actually non-parametric LFs estimated by the classical binned method such as the $1/V_{\mathrm{max}}$
estimator.} of AGNs, many LF models have been proposed. These include pure density evolution \citep[e.g.,][]{1983ApJ...269...35M},
pure luminosity evolution \citep[e.g.,][]{1995ApJ...438..623P,2000MNRAS.317.1014B}
luminosity-dependent density evolution \citep[e.g.,][]{2000A&A...353...25M,2005A&A...441..417H}
luminosity and density evolution \citep[e.g.,][]{2010MNRAS.401.2531A, 2017ApJ...846...78Y},
and luminosity-dependent luminosity evolution \citep[e.g.,][]{2010MNRAS.404..532M,2017MNRAS.469.1912B} models.
The main trend is that LF models are becoming more and more complex, with many non-physical assumptions.
This is a severe drawback for the parametric method.

Non-parametric methods are statistical techniques that make as few assumptions as possible \citep{2001astro.ph.12050W}
about the shape of LF and derive it directly from the sample data.
The explosion of data in future astrophysics provides unique opportunities for non-parametric methods.
With large sample sizes, non-parametric methods make it possible to find subtle effects which might be obscured
by the assumptions built into parametric methods \citep[see][]{2001astro.ph.12050W}. Common non-parametric methods
include various binned methods \citep[e.g.,][]{1968ApJ...151..393S,1980ApJ...235..694A,2000MNRAS.311..433P},
the Step-wise Maximum Likelihood method \citep{1988MNRAS.232..431E}, the \citet{1971MNRAS.155...95L} $C^{-}$ method and its
extended versions \citep[e.g.,][]{1992ApJ...399..345E,1993ApJ...416..450C}.
More recently, some more rigorous statistical techniques have emerged, such as the semi-parametric approach of \citet{2007ApJ...661..703S},
and the Bayesian approach of \citet{2008ApJ...682..874K}. Each of these non-parametric methods has advantages and disadvantages.

Among the non-parametric methods, the binned method is arguably the most popular one, due to its simplicity.
Although more than five decades have passed since its original version \citep[i.e., the $1/V_{\mathrm{max}}$ estimator,][]{1968ApJ...151..393S, 1968MNRAS.138..445R} was presented,
the binned methods are still widely used even in the latest literature
\citep[e.g.,][]{2019A&A...628A...3D,2019A&A...621A.107H,2019MNRAS.488.1035K,2019MNRAS.488.4607L}.
One of the drawbacks of binned methods is that the choice of bin centre and bin width can
significantly distort the shape of the LF, particularly in the
low number density regime. This subsequently impacts on the shape of a
parametric form, which is usually fit to these binned points,
particularly in cases where the $K$-corrections are non-trivial, such as
the case for high-redshift galaxy samples \citep[e.g.][]{2015ApJ...803...34B, 2015MNRAS.452.1817B}.
The $C^{-}$ method and its variants, do not require any binning of data, estimate the cumulative distribution function of the LF.
But they typically assume that luminosity and redshift are statistically independent.
The semi-parametric approach of \citet{2007ApJ...661..703S} is powerful and mathematically rigorous.
But the sophisticated mathematics may make it need time to be recognized and in widespread use.
The Bayesian approach of \citet{2008ApJ...682..874K} can be seen as a combination of parametric and non-parametric methods,
where the LF is modeled as a mixture of Gaussian functions. This approach has many free parameters (typically~$\thicksim22-40$) and
requires critical prior infortation for the parameters.

Motivated by these issues, we have developed a non-parametric method based on kernel density estimation (KDE)
\footnote{The code for performing the KDE analysis in this work is
available upon request from Z. Yuan. A general-purpose code for KDE LF analysis
will be made available on https://github.com/yuanzunli.}.
KDE is a well-established nonparametric approach to estimate continuous density functions based on a
sampled dataset. Due to its effectiveness and flexibility, it has become the most popular method for estimation, interpolation,
and visualization of probability density functions\citep{Botev2010}.
In recent years, KDE is gradually recognized by the astronomical community as an important tool to analyze data
\citep[e.g.,][]{2011A&A...531A.114F,2016MNRAS.459.2618H,2019A&A...630A..55D,2019MNRAS.486.5766T}.
The basic idea of our KDE method is that the contribution of each data point to the LF is regarded as a smooth bump,
whose shape is determined by a Gaussian kernel function $K(z,L)$. Then, KDE sums over all these
bumps to obtain a density estimator. In the analysis, the uncertainties on the measured redshifts and luminosities of
the sources in the sample are ignored.

The paper is organized as follows. Section \ref{intr_kde} gives a brief introduction about the KDE.
Section \ref{LF_KDE} describes the bandwidth selection method for KDE, introduces the boundary bias problem for LF estimate, and specifies the
techniques of estimating LFs by KDE. In Section \ref{result1}, the KDE methods are applied to simulated samples.
Section \ref{result2} shows the result of using our KDE method
to a quasar sample. Section \ref{discussion} gives a comparison of our approach with previous LF estimators.
The main results of the work are summarized in Section \ref{sum}. Throughout the paper, we adopt a Lambda Cold Dark Matter cosmology with the parameters $\Omega_{m}$ = 0.27,  $\Omega_{\Lambda}$ = 0.73, and $H_{0}$ = 71 km s$^{-1}$ Mpc$^{-1}$.

\section{Kernel density estimation}
\label{intr_kde}

\subsection{Definition}
Let $\mathbf{x}=(x_1,x_2,...,x_d)^{T}$ denote a d-dimensional random vector with density $f(\mathbf{x})$
defined on $R^{d}$, and let \{$\boldmath{X}_i,\boldmath{X}_2...,\boldmath{X}_n$\} be a sequence of independent
identically distributed random variables drawn from $f(\mathbf{x})$. The general d-dimensional
multivariate KDE to $f$ is given by
\begin{eqnarray}
\label{kde1}
\hat{f}(\mathbf{x})=\frac{1}{n|\mathbf{H}|^{1/2}}\sum_{j=1}^{n}K(\mathbf{H}^{-1/2}(\mathbf{x}-\boldmath{X}_j)),
\end{eqnarray}
where $K(\cdot)$ is a multivariate kernel function, and $\mathbf{H}$ is the $d\times d$ bandwidth matrix or
smoothing matrix, which is symmetric and positive definite. In this paper, we focus on the 2-dimensional case.

\begin{figure}
  \centerline{
    \includegraphics[scale=0.7,angle=0]{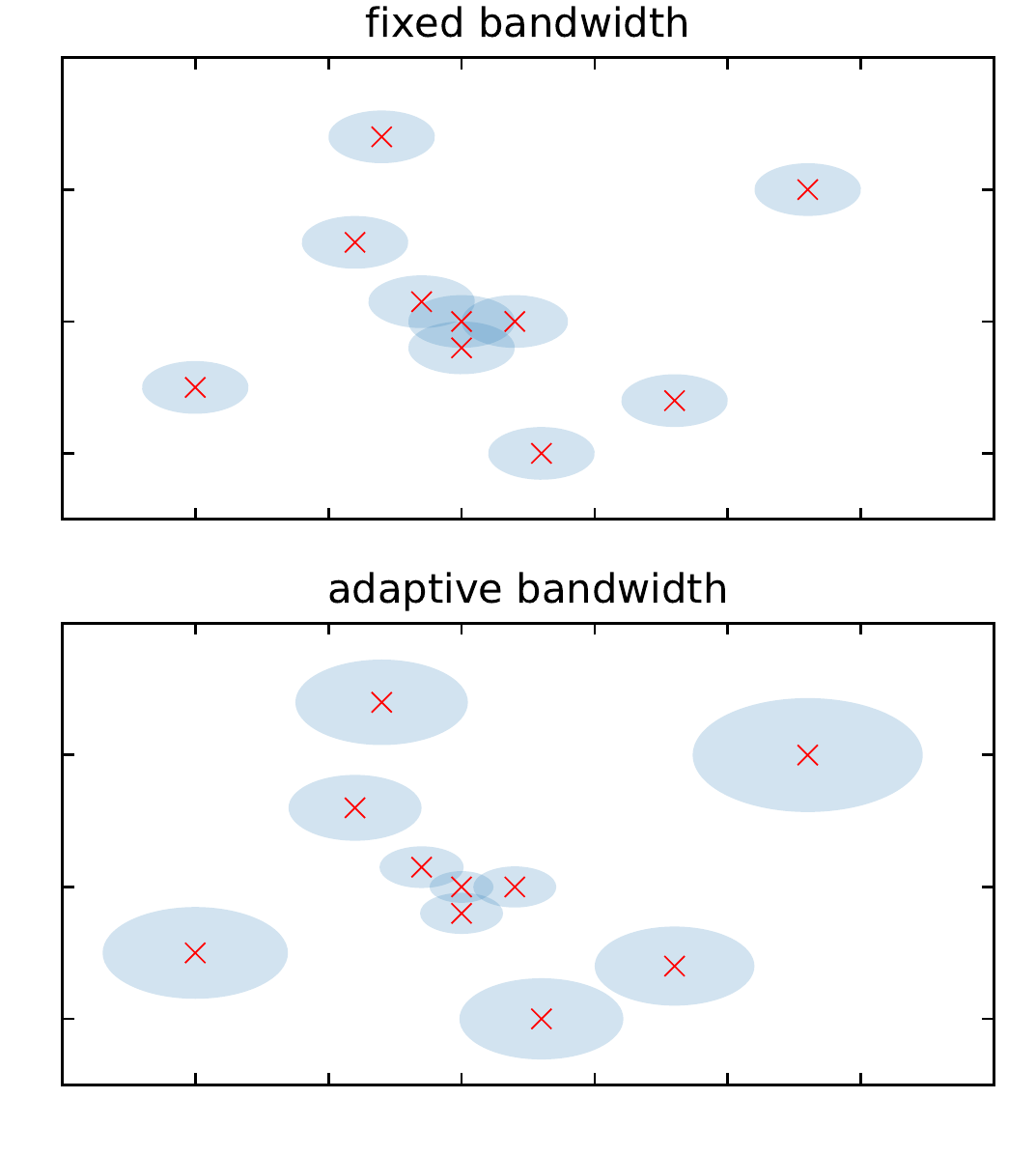}}
  \caption {\label{f1} A toy example demonstrating the idea of the KDE with fixed and adaptive bandwidths.
  Ellipses delineate the kernel spread at one bandwidth away from each hypothetical point (represented by red crosses).}
\end{figure}

KDE can be viewed as a weighted sum of density `bumps'
that are centered at each data point $\boldmath{X}_j$ \citep{Gramacki2018}. The shape of the bumps is determined
by the kernel function $K(\cdot)$, while the width and orientation of the bumps are determined by the bandwidth
matrix $\mathbf{H}$. It is widely accepted that the choice of kernel function is a secondary matter in comparison
with selection of the bandwidth \citep[e.g.,][]{WJ1995,Botev2010,Chen2017,Gramacki2018}.
In most cases, the kernel has the form of a standard multivariate normal density
given by
\begin{eqnarray}
\label{kernel}
K(\mathbf{u})=(2\pi)^{-d/2}\exp(-\frac{1}{2}\mathbf{u}^{T}\mathbf{u}),
\end{eqnarray}
which is what we shall use herein. Choosing the form of $\mathbf{H}$ depends on the complexity of the underlying
density \citep{Sain2002}. Depending on the complexity, $\mathbf{H}$ can be isotropic, diagonal or full matrix, such as
\[
\mathbf{H}=\left[ {\begin{array}{*{20}c}
h^2 & 0   \\
0   & h^2 \\
\end{array}} \right] \mathrm{,}~\mathbf{H}=\left[ {\begin{array}{*{20}c}
h_1^2 & 0 \\
0 & h_2^2 \\
\end{array}} \right] \mathrm{or}~\mathbf{H}=\left[ {\begin{array}{*{20}c}
h_1^2 & h_{12} \\
h_{12} & h_2^2 \\
\end{array}} \right]
\]
The top panel of Figure \ref{f1}  provides a toy example of using the KDE (assuming $\mathbf{H}$ is diagonal) to 10 hypothetical observations.
Ellipses delineate the kernel spread at one bandwidth (the semi axes of each ellipse are $h_1$ and $h_2$) away from each hypothetical point.
If $\mathbf{H}$ is isotropic, these ellipses will degenerate into circles; else if $\mathbf{H}$ is a full matrix,
these ellipses will also have an orientation determined by $h_{12}$.
\citet{Sain2002} argued that for most densities, in particular unimodal ones, allowing different amounts of smoothing
for each dimension (diagonal bandwidth matrix) is adequate. Indeed, for the problem in this work, we find that the choice of
diagonal bandwidth matrix is typically sufficient. If $\mathbf{H}$ is diagonal, Equation (\ref{kde1}) can be simplified to
\begin{eqnarray}
\label{kde2}
\hat{f}(\mathbf{x})=\frac{1}{nh_1h_2}\sum_{j=1}^{n}K(\frac{x_1-X_{1,j}}{h_1},\frac{x_2-X_{2,j}}{h_2}).
\end{eqnarray}

\subsection{Adaptive kernel density estimation}
\label{akde}
Equation (\ref{kde1}) assumes that the bandwidth $\mathbf{H}$ is constant for every individual kernel.
This may lead to poor estimator performance for heterogeneous point patterns \citep{DB2018}.
A popular solution to this problem is using variable bandwidth or adaptive kernel estimator, which allows
$\mathbf{H}$ to vary depending on the local density of the input data points \citep{B1977};
a relatively small bandwidth is needed where observations are densely distributed, and a
large bandwidth is required where observations are sparsely distributed \citep[e.g.,][]{Hu2012}.
As a visual comparison to fixed-bandwidth estimation, the bottom panel of Figure \ref{f1} shows the kernel spread at one
bandwidth away from each hypothetical point for the adaptive estimator.

For the diagonal bandwidth matrix case, the bivariate adaptive kernel estimator \citep[see][]{Sain2002} is given by
\begin{eqnarray}
\label{ada_kde}
\begin{aligned}
\hat{f}_{a}&(\mathbf{x})= \\
&\frac{1}{n}\sum_{j=1}^{n}\frac{1}{h_1(X_j)h_2(X_j)} K(\frac{x_1-X_{1,j}}{h_1(X_j)},\frac{x_2-X_{2,j}}{h_2(X_j)}).
\end{aligned}
\end{eqnarray}
\citet{A1982} suggested that the bandwidths are inversely proportional to the square root of the target density itself.
Obviously, the target density is unknown. According to \citet{S1986}, the practical use of \citet{A1982}'s approach
is implemented as
\begin{eqnarray}
\label{hx}
h_1(\mathbf{x})=h_{10}\tilde{f}(\mathbf{x}|\tilde{h})^{-\beta}, ~\mathrm{and} ~h_2(\mathbf{x})=h_{20}\tilde{f}(\mathbf{x}|\tilde{h})^{-\beta},
\end{eqnarray}
where $\beta \equiv 1/2$, and $\tilde{f}$ is a pilot estimate of the unknown density calculated via Equation (\ref{kde2})
with a fixed pilot bandwidth matrix $\tilde{h}$; $h_{10}$ and $h_{20}$  are overall smoothing parameters for the variable bandwidths
referred to as the global bandwidths \citep{Davies2018}.

\section{Estimating the LF via KDE}
\label{LF_KDE}

\subsection{The luminosity function}

The differential LF of a sample of astrophysical
objects is defined as the number of objects per unit comoving
volume per unit luminosity interval,
\begin{eqnarray}
\label{LFdf}
\phi(z,\mathcal{L})=\frac{d^{2}N}{dVd\mathcal{L}},
\end{eqnarray}
where $z$ denotes redshift and $\mathcal{L}$ denotes the luminosity.
Due the typically large span of the luminosities, it is
often defined in terms of $\log \mathcal{L}$ \citep[e.g.,][]{2008ApJ...686..148C},
\begin{eqnarray}
\label{LFdf}
\phi(z,L)=\frac{d^{2}N}{dVdL},
\end{eqnarray}
where $L\equiv \log \mathcal{L}$ denotes the logarithm of luminosity.

In an actual survey, only a very limited number of objects in
the universe can be observed, as our survey only
covers a fraction of the sky and is subject to a selection function.
Thus the estimation of LFs is inevitably based on a truncated
sample of objects. The integral of the LF over the survey region $W$ should approximate to the sample size $n$,
for sufficiently large n, i.e.,
\begin{eqnarray}
\label{eqint}
\int\!\!\!\!\int_{W}\phi(z,L)\Omega\frac{dV}{dz}dzdL=n,
\end{eqnarray}
where $\Omega$ is the solid angle subtended by the survey, and $dV/dz$ is the differential comoving volume per unit solid angle \citep{1999astro.ph..5116H}.
The LF is related to the probability distribution of $(z,L)$ by
\begin{eqnarray}
\label{p_phi}
p(z,L)=\frac{\Omega}{n}\phi(z,L)\frac{dV}{dz}.
\end{eqnarray}
Once we have an estimate of $p(z,L)$ by the KDE, we can easily convert this to an
estimate of $\phi(L, z)$ using Equation (\ref{p_phi}).
It needs to be emphasized that the domain of $p(z,L)$ is the survey region $W$, thus
the domain of the estimated LF by Equation (\ref{p_phi}) is also limited to $W$.

\subsection{Optimal bandwidth selection for KDE}
Choosing the bandwidth parameter is the most crucial issue in using KDE to estimate $p(z,L)$,
as the accuracy of KDE depends very strongly on the bandwidth matrix \citep[e.g.,][]{Gramacki2018}.
Basically, all the common approaches to bandwidth selection are
based on some appropriate error criteria, which measures the closeness of $\hat{f}$ to its target density $f$.
One of the most well-known error measurements is the mean integrated square error (MISE). The optimal bandwidth can
be obtained by minimizing the Asymptotic MISE (AMISE), the dominating quantity in the MISE \citep{Chen2017}.
The three major types of bandwidth selectors include the rule of thumb, cross-validation (CV) methods, and plug-in method.
In this work, we focus on the CV method and our selector is based on the likelihood cross-validation (LCV) criterion.

\subsubsection{Likelihood cross-validation}
The LCV thinks about the estimated density itself as a likelihood function, and the LCV objective function is given by
\begin{eqnarray}
\label{lcv}
\mathrm{LCV}(\mathbf{H}|\boldmath{X})=\frac{1}{n}\sum_{i=1}^{n}\log \hat{f}_{-i}(\boldmath{X}_i),
\end{eqnarray}
where $\hat{f}_{-i}$ is the leave-one-out estimator
\begin{eqnarray}
\label{fi1}
\hat{f}_{-i}(\boldmath{X}_i)=\!\!\frac{1}{n-1}\!\sum_{j=1 \atop j\neq i}^{n} \!|\mathbf{H}|^{-1/2}K(\mathbf{H}^{-1/2}(\boldmath{X}_i \!\! - \!\! \boldmath{X}_j)).
\end{eqnarray}
If $\mathbf{H}$ is diagonal, Equation (\ref{fi1}) can be simplified to
\begin{eqnarray}
\label{fi2}
\begin{aligned}
\hat{f}_{-i}(X_i)&=\frac{1}{n-1} \times \\
&\sum_{j=1 \atop j\neq i}^{n} \frac{1}{h_1h_2}K(\frac{X_{1,i}-X_{1,j}}{h_1},\frac{X_{2,i}-X_{2,j}}{h_2}).
\end{aligned}
\end{eqnarray}
The LCV bandwidth matrix $\hat{\mathbf{H}}_\mathrm{LCV}$ is the maximizer of the LCV(H) objective function \citep[e.g.,][]{Davies2018}
\begin{eqnarray}
\label{Hlcv}
\hat{\mathbf{H}}_\mathrm{LCV}=\mathrm{argmax}[\mathrm{LCV}(\mathbf{H}|\boldmath{X})]
\end{eqnarray}
and it has to be maximized numerically. In view of the difficult numerical procedure as the dimension of data increases, \citet{Zhang2006}
improved the LCV method from a Bayesian perspective. They treated non-zero components of $\mathbf{H}$ as parameters, whose posterior density
can be obtained by the Markov Chain Monte Carlo (MCMC) technique based on the LCV criterion. Given $\mathbf{H}$, the logarithmic likelihood function of
\{$\boldmath{X}_i,\boldmath{X}_2...,\boldmath{X}_n$\} is
\begin{eqnarray}
\label{loglik}
L(\boldmath{X}_i,\boldmath{X}_2...,\boldmath{X}_n|\mathbf{H})=\sum_{i=1}^{n} \log \hat{f}_{-i}(\boldmath{X}_i).
\end{eqnarray}
However, as pointed by \citet{Zhang2006}, when using the likelihood given by Equation (\ref{loglik}) to perform the Bayesian inference,
sufficient prior information on components of $\mathbf{H}$ is
required. Using this likelihood function in our problem does not lead to convergence
for the MCMC algorithm, we therefore consider a new likelihood function.

\subsubsection{A new likelihood cross-validation method}

The new candidate likelihood function derives from \citet{1983ApJ...269...35M}.
In their analysis the probability distribution in the likelihood
for the observables is described by Poisson probabilities. The $L-z$ space is divided into small intervals of size $dLdz$.
In each element, the expected number of objects at $z$, $L$ is
$\lambda(z,L)dzdL=\Omega\phi(z,L)(dV/dz)dzdL$. The likelihood function, $\mathscr{L}$,
is defined as the product of the probabilities of observing one object at each ($z_i,L_i$)
element and the probabilities of observing zero object in all other ($z_j,L_j$) elements
in the accessible regions of the $z-L$ plane \citep{1983ApJ...269...35M}.
Using Poisson probabilities, one has
\begin{eqnarray}
\label{loglik}
\mathscr{L}\!=\!\!\!\prod_{i}^{n}\! [\lambda(z_i, \! L_i)dzdL e^{-\lambda(z_i, \! L_i)dzdL}]\! \prod_j \! e^{-\lambda(z_j,\! L_j)dzdL}
\end{eqnarray}
Transforming to the standard expression $S\equiv-2 \ln \mathscr{L }$ and dropping terms which
are not model dependent, we obtain
\begin{eqnarray}
\label{likelihood1}
\begin{aligned}
S=-2\!\!\sum_{i}^{n}\!\ln[\phi(z_{i},L_{i})]+2\!\!\!\int\!\!\!\!\int_{W}\!\!\!\phi(z,L)\Omega\frac{dV}{dz}dzdL.
\end{aligned}
\end{eqnarray}
The likelihood function given above naturally incorporates the selection function of the sample
(by considering the integral interval $W$), and it has proved popular \citep{2011A&ARv..19...41J} and has been widely applied
in parametric estimation of LFs \citep[e.g.,][]{2012ApJ...751..108A,2016ApJ...829...95Y,2019MNRAS.488.1035K}.

Inserting Equation~(\ref{p_phi}) into Equation~(\ref{likelihood1}) and dropping terms which are not model dependent, we obtain
\begin{eqnarray}
\label{likelihood2}
\begin{aligned}
S=-2\!\!\sum_{i}^{n}\!\ln[p(z_{i},L_{i})]+2n\!\!\int\!\!\!\!\int_{W}\!\!\!p(z,L)dzdL.
\end{aligned}
\end{eqnarray}
Equation (\ref{likelihood2}) enables one to transform the LF measurement into a
problem of probability density estimation. $p(z,L)$ can be replaced with its KDE form $\hat{p}(z,L|h_1,h_2)$ given by
Equation (\ref{kde2}), and $p(z_i,L_i)$ is replaced with the leave-one-out estimator $\hat{p}_{-i}(z_i,L_i|h_1,h_2)$
constructed by Equation (\ref{fi2}). Then we have
\begin{eqnarray}
\label{likelihood3}
\begin{aligned}
S=-2\sum_{i}^{n}&\ln[\hat{p}_{-i}(z_{i},L_{i}|h_1,h_2)] ~ + \\
&2n\!\!\int\!\!\!\!\int_{W}\hat{p}(z,L|h_1,h_2)dzdL.
\end{aligned}
\end{eqnarray}
Note that in the first term $\hat{p}_{-i}$ should not be confused with $\hat{p}$. The term `leave-one-out' means
that the observations used to calculate $\hat{p}_{-i}(z_i,L_i)$ are independent of $(z_i,L_i)$. This is where the name
`cross-validation' comes from: using one subset of data to make analysis on another subset \citep[see][]{Gramacki2018}.

Using Equation (\ref{likelihood3}), we can obtain the optimal bandwidth by minimizing $S$
\begin{eqnarray}
\label{Hs}
\hat{\mathbf{H}}_S=\mathrm{argmin}[\boldmath{S}(\mathbf{H}|\boldmath{z},\boldmath{L})].
\end{eqnarray}
We find the above bandwidth selector performs better than the LCV selector given by Equation (\ref{Hlcv}).
We note that the numerical optimization result of the LCV selector depends on the initial values, suggesting
it may have more than one local maximum in the objective function. In this respect, our new selector has better stability.

Equation~(\ref{likelihood3}) can also be used to perform Bayesian
inference by combining it with prior information on the
bandwidths $h_1$ and $h_2$. We find that using uniform priors for $h_1$ and $h_2$ are sufficient to employ the MCMC algorithm.
Therefore our new likelihood function is a better choice than the one given by Equation (\ref{loglik}), that requires
more critical prior information.

\begin{figure*}
  \centerline{
    \includegraphics[scale=0.65,angle=0]{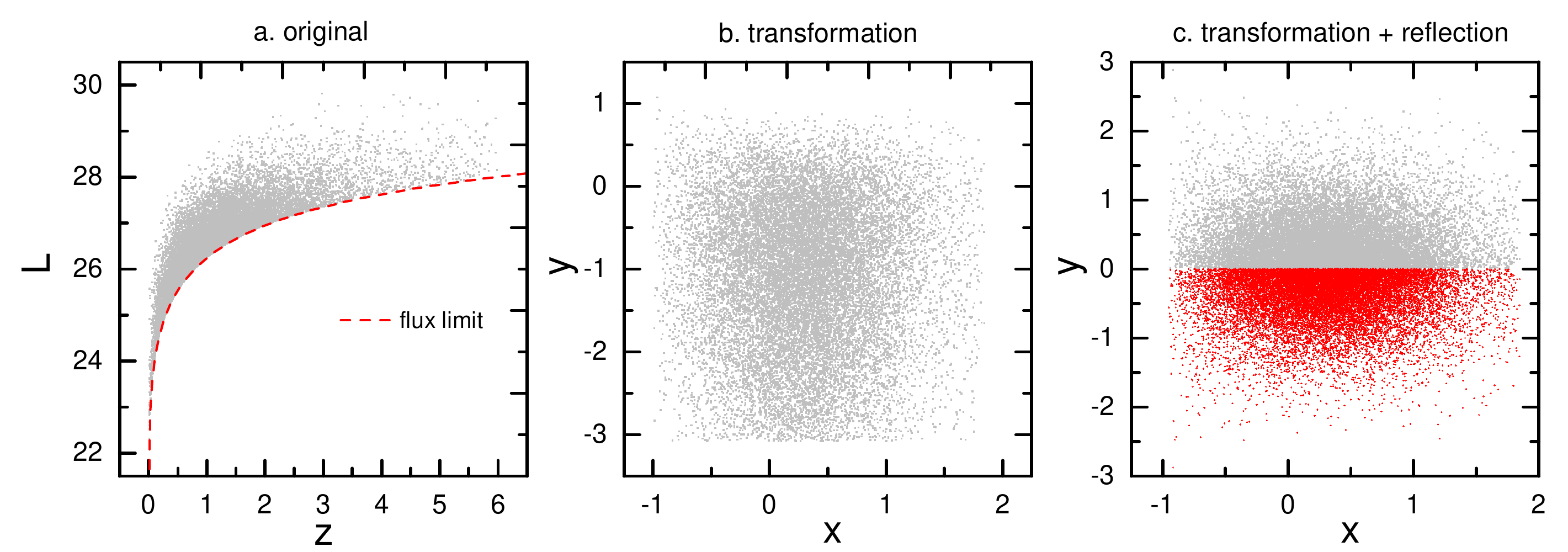}}
  \caption {\label{f2} Panel $\mathbf{a}$: a simulated sample illustrates the truncation bound (red dashed line). More details about the simulation are given in Section \ref{result1}.
  Panel $\mathbf{b}$: This shows how the simulated data in panel $\mathbf{a}$ will look like after transformation by Equation (\ref{trans}). Panel $\mathbf{c}$: The light gray points show the
  transformed data according to Equation (\ref{trans2}), and the red points are reflection points.}
\label{RLF}
\end{figure*}

\subsection{Boundary effects for LF estimate}
\label{truncation_boundary}
Recall that the domain of $p(z,L)$ in Equation (\ref{p_phi}) is the survey region $W$.
This means that the estimate of $p(z,L)$ using KDE is
based on bounded data, while certain difficulties can arise at the boundaries and near them, known as
boundary effects or boundary bias \citep[e.g.,][]{M1999}.

In astronomy, many surveys are flux-limited. When the observation points are plotted on the $L-z$ plane,
there is a clear truncation bound (see Figure \ref{f2} a) defined by $f_{\mathrm{lim}}(z)$:
\begin{eqnarray}
\label{f_lim}
f_{\mathrm{lim}}(z)=4\pi d_L^{2}(z)(1/K(z))F_{\mathrm{lim}},
\end{eqnarray}
where $d_L(z)$ is the luminosity distance, $F_{\mathrm{lim}}$ is the survey flux limit, and $K(z)$ represents the $K$-correction.
One has $K(z)=(1+z)^{1-\alpha}$ for a power-law emission spectrum of index $\alpha$ \citep[e.g.,][]{2011ApJ...743..104S}.
To give reliable results, any LF estimator must treat the truncation
boundary, or flux-limit of the survey, properly. From Equation~(\ref{f_lim}),
we can see that only when $K(z)=(1+z)^{1-\alpha}$ and $\alpha$ is constant, the truncation bound
defined by $f_{\mathrm{lim}}(z)$ is a curve. Otherwise, if the $K$-correction $K(z)$ takes a more complex form, the truncation bound
will be a 2-dimensional region but not a simple curve. In this
section, we only consider the simple case for the truncation
boundary is a simple curve.

The mathematical description for the above is that, suppose we observe $n$ points in a 2-dimensional space, $\{(z_i,L_i), i=1,2,...,n\}$,
these points are always from a bounded subset $W$ of the plane; $(\boldmath{z},\boldmath{L}) \in W \subset \mathbf{R}^2$,
and $W$ is referred to as \emph{study window} \citep[e.g.,][]{Davies2018} or \emph{survey region}. A direct use of the KDE formula
to $W$ may lead to the problem of underestimating the density at boundaries.
The reason for the boundary problem is that the kernels from data points
near the boundary lose their full probability weight and points that
lie just outside the boundary have
no opportunity to contribute to the final density estimation on $W$ \citep[e.g.,][]{Davies2018}.
The boundary effect has long been recognized \citep{G1979}, and it is an import research topic in KDE study
\citep[e.g.,][]{MR1994,HP2002,MH2010,MS2014}. The common techniques for reducing boundary effects include, reflection of
data \citep[e.g.,][]{Jones1993}, transformation of data \citep[e.g.,][]{MR1994}, and boundary kernel estimators \citep[e.g.,][]{HP2002}.
For the LF estimate problems, we find that the transformation method, and a method of combining transformation and reflection \citep{KA2005}
are able to reduce boundary effects.

\subsubsection{The transformation method}
\label{TM}

The basic idea is that, take a one-to-one and continuous function $g$, and use the regular kernel estimator with the transformed
data set $\{g(X_1),g(X_2),...,g(X_n)\}$. For the LF estimate, we use the following transformation:
\begin{eqnarray}
\label{trans}
x=\ln(z+\delta_1), ~\mathrm{and}, ~ y=\ln(L-f_{\mathrm{lim}}(z)+\delta_2),
\end{eqnarray}
where $f_{\mathrm{lim}}(z)$ is given by Equation~(\ref{f_lim}), and $\delta_1$ and $\delta_2$ are transformation parameters.
Figure \ref{f2}b provides a toy example showing how a typical
flux-limited data set will look like after transformation by Equation~(\ref{trans}).

The Jacobian matrix for the above transformation is
\[
\mathbf{J}=\left[ {\begin{array}{*{20}l}
     \frac{\partial x}{\partial z}=\frac{1}{z+\delta_1} & \frac{\partial x}{\partial L} =0 \\
     \frac{\partial y}{\partial z}=\frac{-f_{\mathrm{lim}}'(z)}{L-f_{\mathrm{lim}}(z)+\delta_2} & \frac{\partial y}{\partial L} =\frac{1}{L-f_{\mathrm{lim}}(z)+\delta_2}\\
\end{array}} \right],
\]
and $\mathrm{det}(\mathbf{J})=|\mathbf{J}|=\frac{1}{(z+\delta_1)(L-f_{\mathrm{lim}}(z)+\delta_2)}$.
After the transformation, the density of $(x,y)$, denoted as $\hat{f}_{\mathrm{t}}(x,y)$, can be estimated by Equation (\ref{kde2}),
and its leave-one-out estimator, $\hat{f}_{\mathrm{t},-i}(x_i,y_i)$, is constructed by Equation (\ref{fi2}).
Then by transforming back to the density of original data set, we have
\begin{eqnarray}
\label{ppi}
\begin{cases}
  \displaystyle \hat{p}_{\mathrm{t}}(z,L|h_1,h_2,\delta_1,\delta_2)=\frac{\hat{f}_{\mathrm{t}}(x,y)}{(z \! + \! \delta_1)(L \!- \!f_{\mathrm{lim}}(z)\!+\! \delta_2)}, \\
 \displaystyle  \hat{p}_{\mathrm{t},-i}(z_i,L_i|h_1,h_2,\delta_1,\delta_2)=\frac{\hat{f}_{\mathrm{t},-i}(x_i,y_i)}{(z_i \!+\!\delta_1)(L_i\!-\!f_{\mathrm{lim}}(z_i)\!+\!\delta_2)}.
\end{cases}
\end{eqnarray}
Inserting Equation~(\ref{ppi}) into (\ref{likelihood3}), we obtain the negative logarithmic likelihood function $S$.
Following \citet{Liu2011}, we estimate the optimal bandwidths and transformation parameters $\delta_1$ and $\delta_2$ simultaneously.
They can be estimated by numerically minimizing the object function
$S$, or by combining with uniform priors
for $h_1$, $h_2$, $\delta_1$ and $\delta_2$, one can also employ the MCMC algorithm to sample the bandwidth and
transformation parameters simultaneously. The MCMC algorithm used in this work is ``CosmoMC'',
a public Fortran code of \citet{2002PhRvD..66j3511L}.
Once we know the probability density $\hat{p_{\mathrm{t}}}$, the KDE of the LF is
easily obtained by
\begin{eqnarray}
\label{phi_t}
\hat{\phi}_{\mathrm{t}}(z,L)=\hat{p}_{\mathrm{t}}(z,L|h_1,h_2,\delta_1,\delta_2)n (\Omega\frac{dV}{dz})^{-1}.
\end{eqnarray}

\subsubsection{The transformation-reflection method}
\label{trm}
Introduced by \citet{KA2005}, the transformation-reflection method was originally used for univariate data.
We extend the method to the bivariate case, and extending it to the
trivariate case is also possible (see section \ref{tr_3d}). First, we transform the original data by
\begin{eqnarray}
\label{trans2}
x=\ln(z+\delta_1), ~\mathrm{and}, ~ y=L-f_{\mathrm{lim}}(z),
\end{eqnarray}
where the meanings of $f_{\mathrm{lim}}$ and $\delta_1$ are similar to those in Equation~(\ref{trans}).
The determinant of the Jacobian matrix for the above transformation is $\mathrm{det}(\mathbf{J})=\frac{1}{(z+\delta_1)}$.
Second, we add the missing `probability mass' \citep[e.g.,][]{Gramacki2018} represented by the data
set $\{(x_1,-y_1),(x_2,-y_2),...,(x_n,-y_n)\}$.
For illustration of the above ideas, Figure \ref{f1}$\mathbf{c}$ shows
how a typical flux-limited data set
will look after transformation and reflection.
The KDE to the density of $(x,y)$ is
\begin{eqnarray}
\label{kde_r}
\begin{aligned}
\hat{f}_{\mathrm{tr}}(x,y&)=\frac{2}{2nh_1h_2} \times \\
&\sum_{j=1}^{n} \left( K(\frac{x\!-\!x_j}{h_1},\frac{y\!-\!y_j}{h_2}) \!+\! K(\frac{x\!-\!x_j}{h_1},\frac{y\!+\!y_j}{h_2}) \right ).
\end{aligned}
\end{eqnarray}
The corresponding leave-one-out estimator is
\begin{eqnarray}
\label{kde_ri}
\begin{aligned}
&\hat{f}_{\mathrm{tr},-i}(x_i,y_i)=\frac{2}{(2n-1)h_1h_2} \times \\
&\left(\! \sum_{j=1 \atop j\neq i}^{n} \! K(\frac{x_i\!-\!x_j}{h_1},\frac{y_i\!-\!y_j}{h_2}) \!+\! \sum_{j=1}^{n} \!K(\frac{x_i\!-\!x_j}{h_1},\frac{y_i\!+\!y_j}{h_2}) \!\right).
\end{aligned}
\end{eqnarray}
Then by transforming back to the density of original data set, we have
\begin{eqnarray}
\label{pp_tr}
\begin{cases}
  \displaystyle \hat{p}_{\mathrm{tr}}(z,L|h_1,h_2,\delta_1)=\frac{\hat{f}_{\mathrm{tr}}(x,y)}{(z \! + \! \delta_1)}, \\
 \displaystyle  \hat{p}_{\mathrm{tr},-i}(z_i,L_i|h_1,h_2,\delta_1)=\frac{\hat{f}_{\mathrm{tr},-i}(x_i,y_i)}{(z_i \!+\!\delta_1)}.
\end{cases}
\end{eqnarray}
Inserting Equation (\ref{pp_tr}) into (\ref{likelihood3}), we can obtain the likelihood function for the transformation-reflection estimator.
The next steps for estimating the bandwidth and transformation
parameters are similar to those described in Section~\ref{TM}. Finally,
we can obtain the LF estimated by the transformation-reflection approach,
\begin{eqnarray}
\label{phi_tr}
\hat{\phi}_{\mathrm{tr}}(z,L)=\hat{p}_{\mathrm{tr}}(z,L|h_1,h_2,\delta_1)n (\Omega\frac{dV}{dz})^{-1}.
\end{eqnarray}

\subsubsection{The transformation-reflection adaptive KDE approach}
\label{adaptive_kde}

In Equation (\ref{kde_r}), the bandwidths $h_1$ and $h_2$ are constant for every individual kernel.
A useful improvement is to use variable bandwidths depending on the local density of the input data
points. This can be achieved using the transformation-reflection adaptive KDE ($\hat{\phi}_{\mathrm{tra}}$) approach.
The $\hat{\phi}_{\mathrm{tra}}$ estimator is implemented on the basis of $\hat{\phi}_{\mathrm{tr}}$, and it involves the following steps:

\begin{enumerate}
  \item Employ the transformation-reflection method, and obtain the optimal bandwidths and the transformation parameter,
  denoted as $\tilde{h}_1$, $\tilde{h}_2$, and $\tilde{\delta}_1$.

  \item Calculate $\tilde{f}_{\mathrm{tr}}(x,y)$ via Equation (\ref{kde_r}), given $h_1=\tilde{h}_1$, $h_2=\tilde{h}_2$, and $\delta_1=\tilde{\delta}_1$.

  \item Let the bandwidths vary with the local density:
\begin{eqnarray}
\label{h1h2x}
\begin{cases}
  \displaystyle  h_1(x,y)=h_{10}\tilde{f}_{\mathrm{tr}}(x,y| \tilde{h}_1, \tilde{h}_2, \tilde{\delta}_1)^{-\beta}\\
 \displaystyle   h_2(x,y)=h_{20}\tilde{f}_{\mathrm{tr}}(x,y| \tilde{h}_1, \tilde{h}_2, \tilde{\delta}_1)^{-\beta},
\end{cases}
\end{eqnarray}
where $h_{10}$ and $h_{20}$ are global bandwidths which need to be determined. Unlike Equation (\ref{hx}) where $\beta \equiv 1/2$,
we let $\beta$ be a free parameter here.

  \item In Equations (\ref{kde_r}) and (\ref{kde_ri}), replace $h_1$ and $h_2$ with $h_1(x,y)$ and $h_2(x,y)$, respectively.
 Then we can obtain the adaptive KDE $\hat{f}_{\mathrm{tra}}(x,y)$ and its leave-one-out estimator $\hat{f}_{\mathrm{tra,-i}}(x,y)$.

 \item Determine $h_{10}$, $h_{20}$ and $\beta$ via the maximum likelihood method or the MCMC algorithm. The detailed formulas for constructing
 $\hat{f}_{\mathrm{tra}}(x,y)$ and $\hat{f}_{\mathrm{tra,-i}}(x,y)$, as well as the process for determining their parameters can be found in
 Appendix \ref{A1}.
\end{enumerate}

\begin{figure*}
  \centerline{
    \includegraphics[scale=0.55,angle=0]{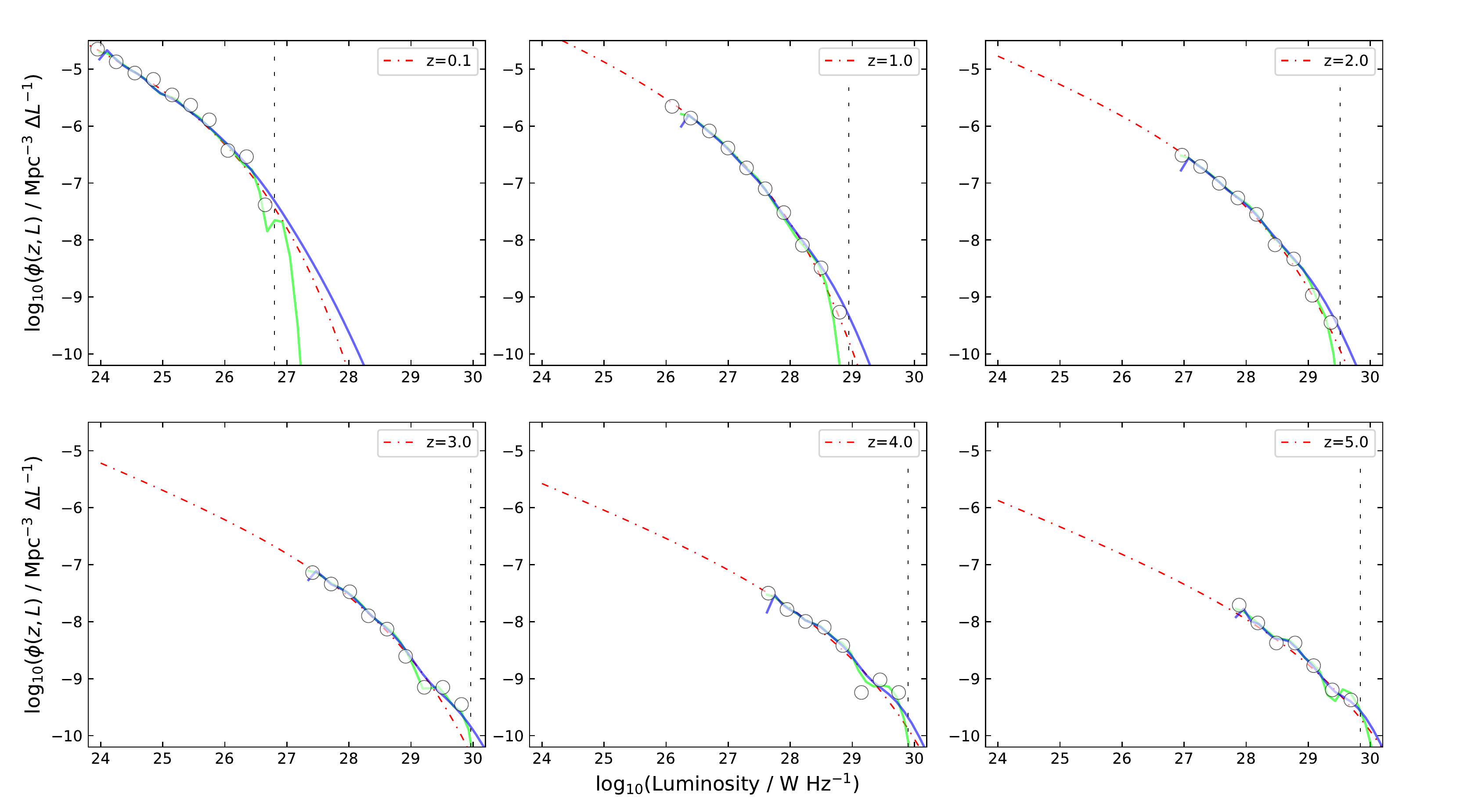}}
  \caption {\label{fig_ref} RLFs estimated by the $\hat{\phi}_{\mathrm{t}}$ (blue solid lines) and $\hat{\phi}_{\mathrm{tr}}$ (green solid lines) estimators at several redshifts,
  compared with the estimates (black open circles) given by the binned method of \citet{2000MNRAS.311..433P}. The red dash-dotted lines represent the true RLF.
  The vertical dashed lines mark the higher luminosity limits of the simulated survey at different redshifts.}
\end{figure*}

\begin{figure}
  \centerline{
    \includegraphics[scale=0.5,angle=0]{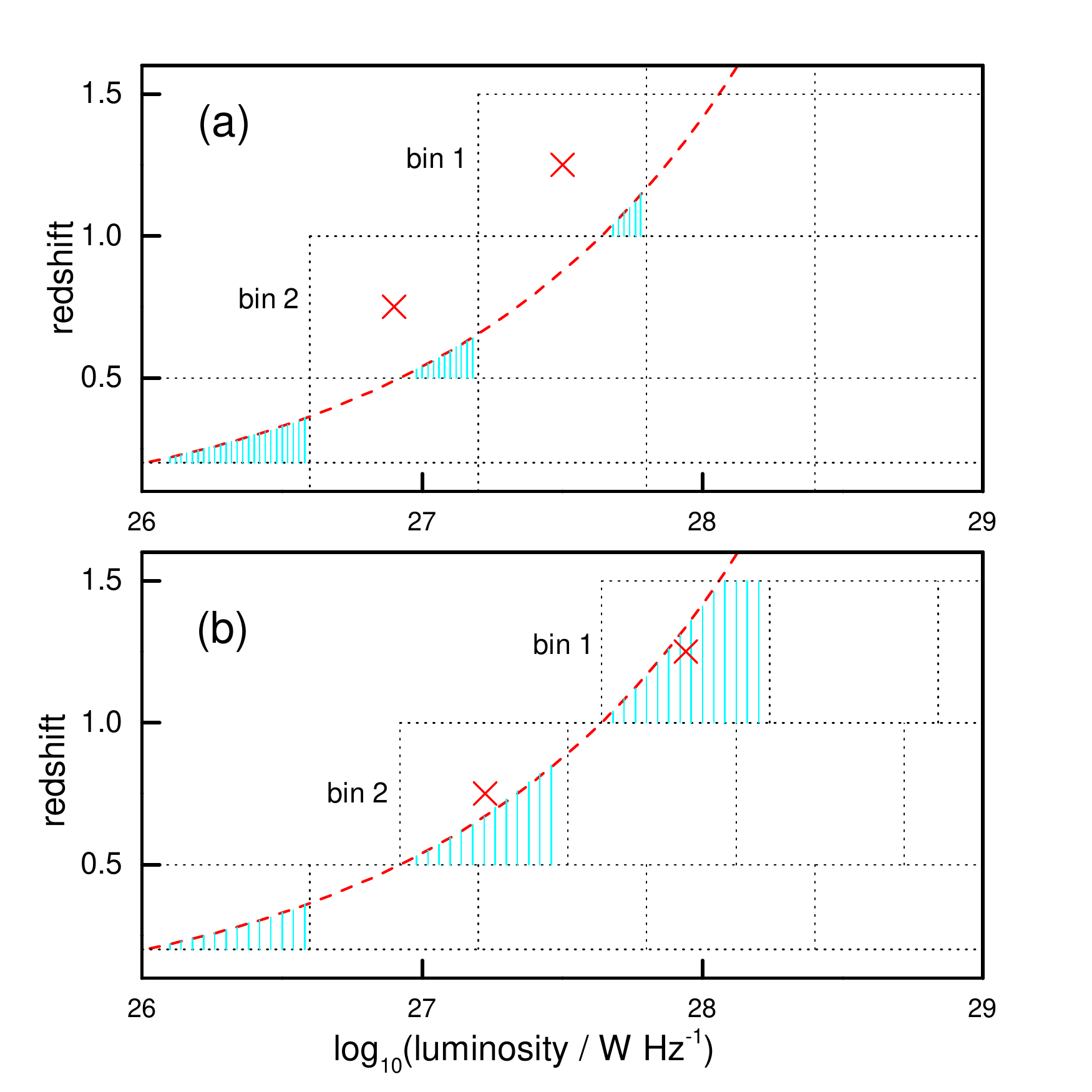}}
  \caption {\label{bin} A toy example illustrates different schemes for dividing bins. In each panel the red dashed line shows the
  flux limit curve, red crosses mark the centers of bins 1 and 2, and the shaded regions represent the surveyed regions in the bins.
  Panel (a): both the redshift and luminosity intervals are chosen arbitrarily. This may lead to the bins located at the faint end
  (e.g., bins 1 and 2) to contain very few objects and cause biases with small number statistics.
  More seriously, the binned estimator typically reports the density of a bin center as its result, but for bins 1 and 2,
  the bin centers are far away from the surveyed regions. In this situation, the bined LF at the faint end would be significantly biased.
  Panel (b): the redshift bins are chosen arbitrarily, while the starting positions of luminosity bins are determined by the intersecting points of
  the flux limit curve and redshift grid-lines. This dividing scheme can alleviate the faint end bias \citep[see][for details]{2013ApSS.345..305Y}.}
\end{figure}

\begin{figure*}[!htb]
\includegraphics[height=11cm,width=11cm]{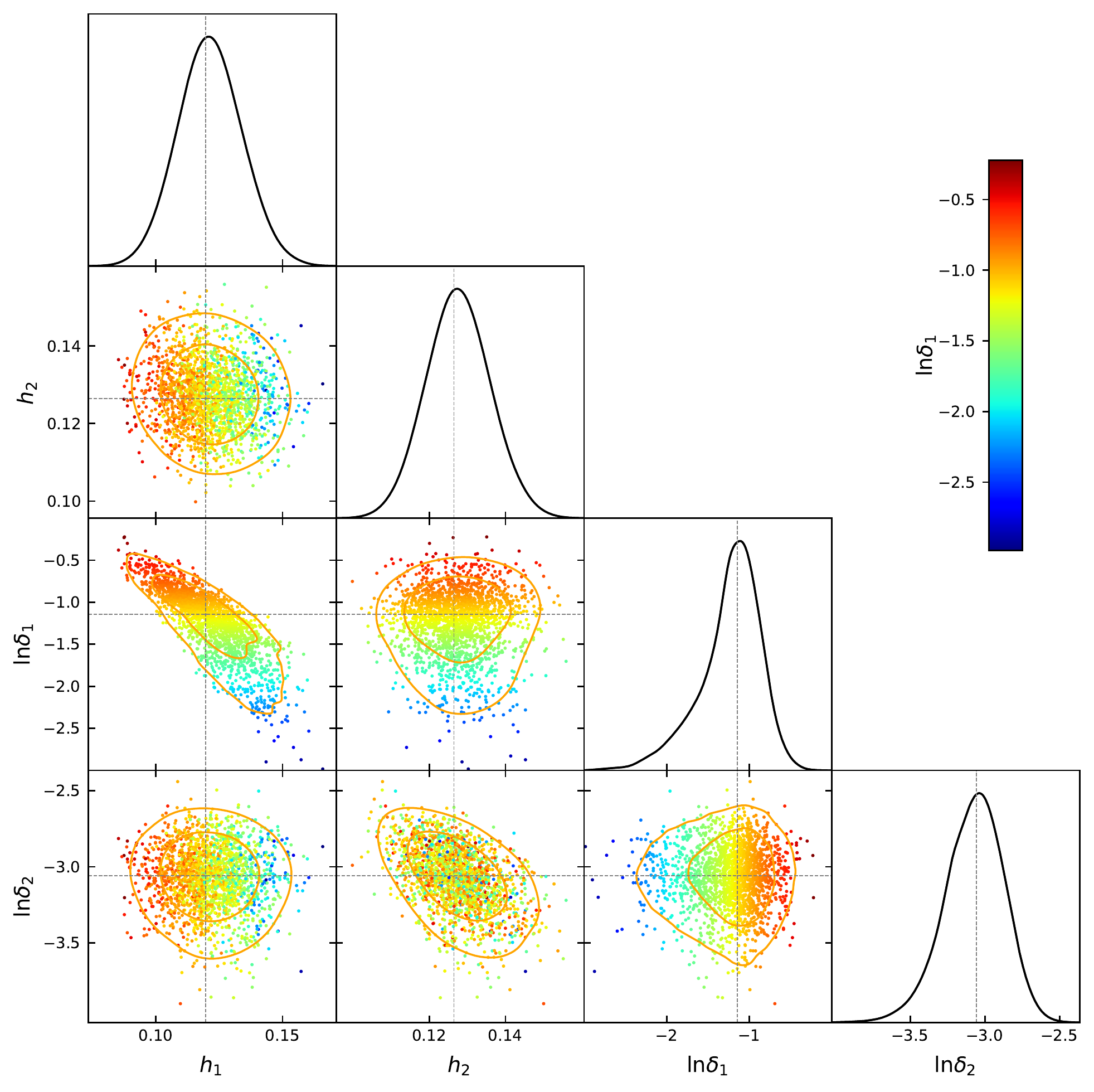}
\\[18pt]
\includegraphics[height=8cm,width=8.25cm]{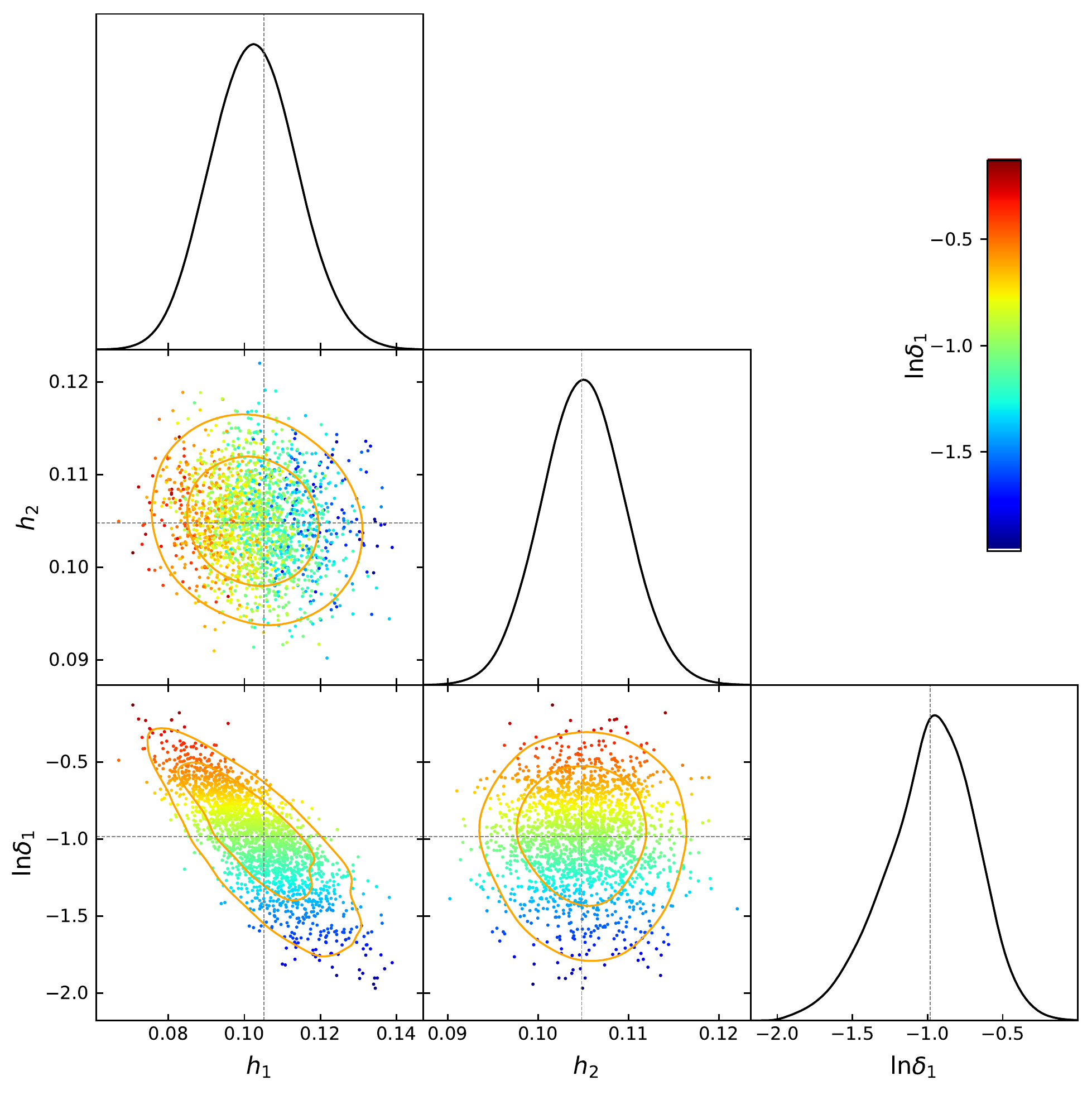}  \qquad
\includegraphics[height=8cm,width=8.25cm]{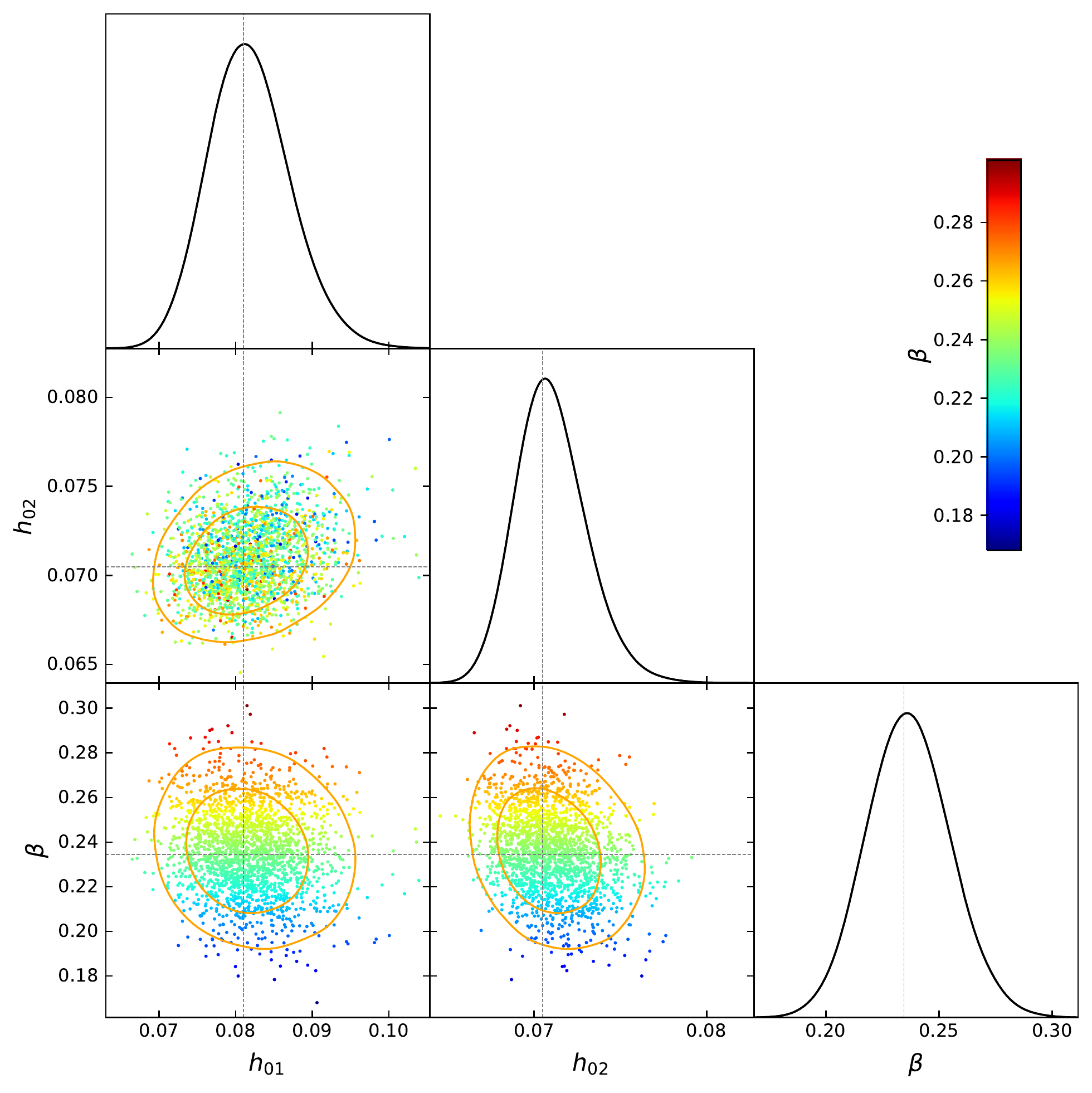}
\caption{Marginalized 1D and 2D posterior distributions of parameters in the KDE methods. The contours containing 68\% and 95\% of the probability.
MCMC samples are shown as coloured points, where the colour corresponds to the parameter shown in the colour bar. The vertical lines mark the locations
of best-fit values by MCMC. The upper, lower left, and lower right panels correspond to $\hat{\phi}_{\mathrm{t}}$,
$\hat{\phi}_{\mathrm{tr}}$ and $\hat{\phi}_{\mathrm{tra}}$ respectively. The triangular plots are created by the python package GetDist \citep{2019arXiv191013970L}.}
\label{fig_mcmc}
\end{figure*}

\begin{figure*}
  \centerline{
    \includegraphics[scale=0.55,angle=0]{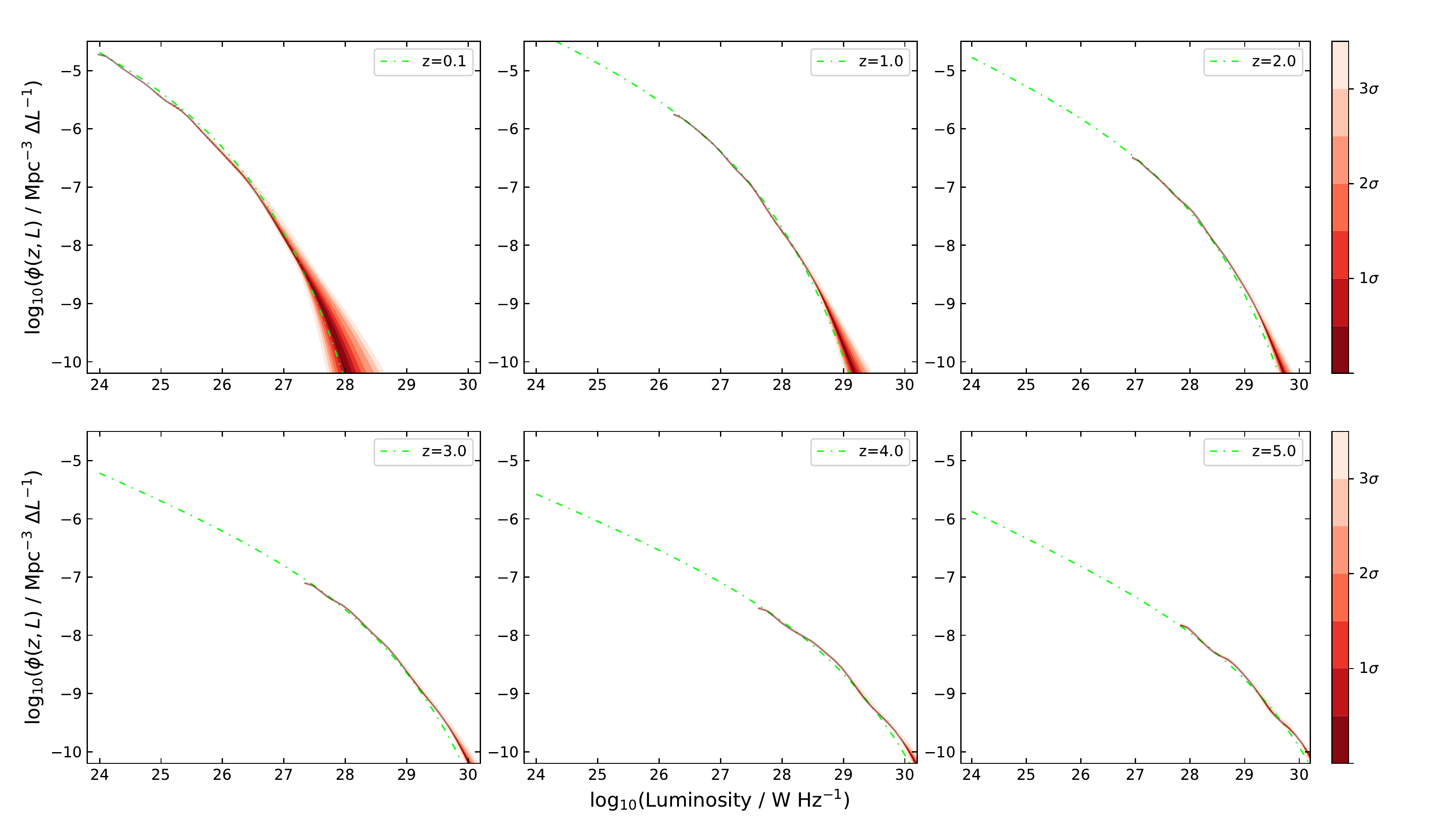}}
  \caption {\label{fig_ada} The RLF and its uncertainty estimated using the $\hat{\phi}_{\mathrm{tra}}$ at several redshifts for the simulated sample. The green dash-dotted lines represent the true RLF.}
\end{figure*}

\begin{figure*}
  \centerline{
    \includegraphics[scale=0.55,angle=0]{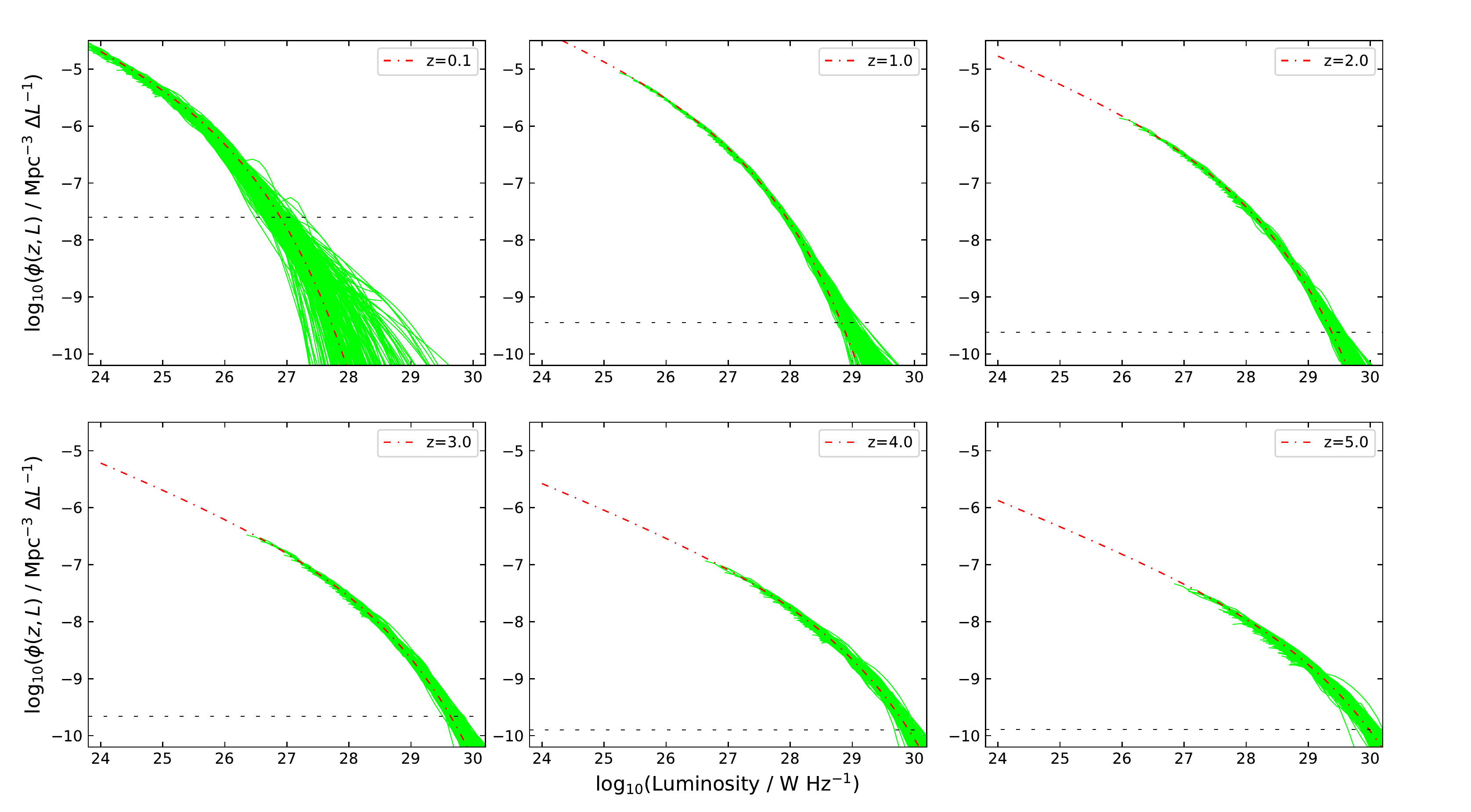}}
  \caption {\label{f7} RLFs (green lines) estimated using $\hat{\phi}_{\mathrm{tra}}$ for 200 simulated samples at several redshifts. The red dash-dotted lines represent the true RLF.
  The horizontal dashed lines mark the approximate observational limits of the simulated surveys, below which the RLFs are extrapolated by $\hat{\phi}_{\mathrm{tra}}$.}
\label{RLF}
\end{figure*}

\begin{figure*}
  \centerline{
    \includegraphics[scale=0.55,angle=0]{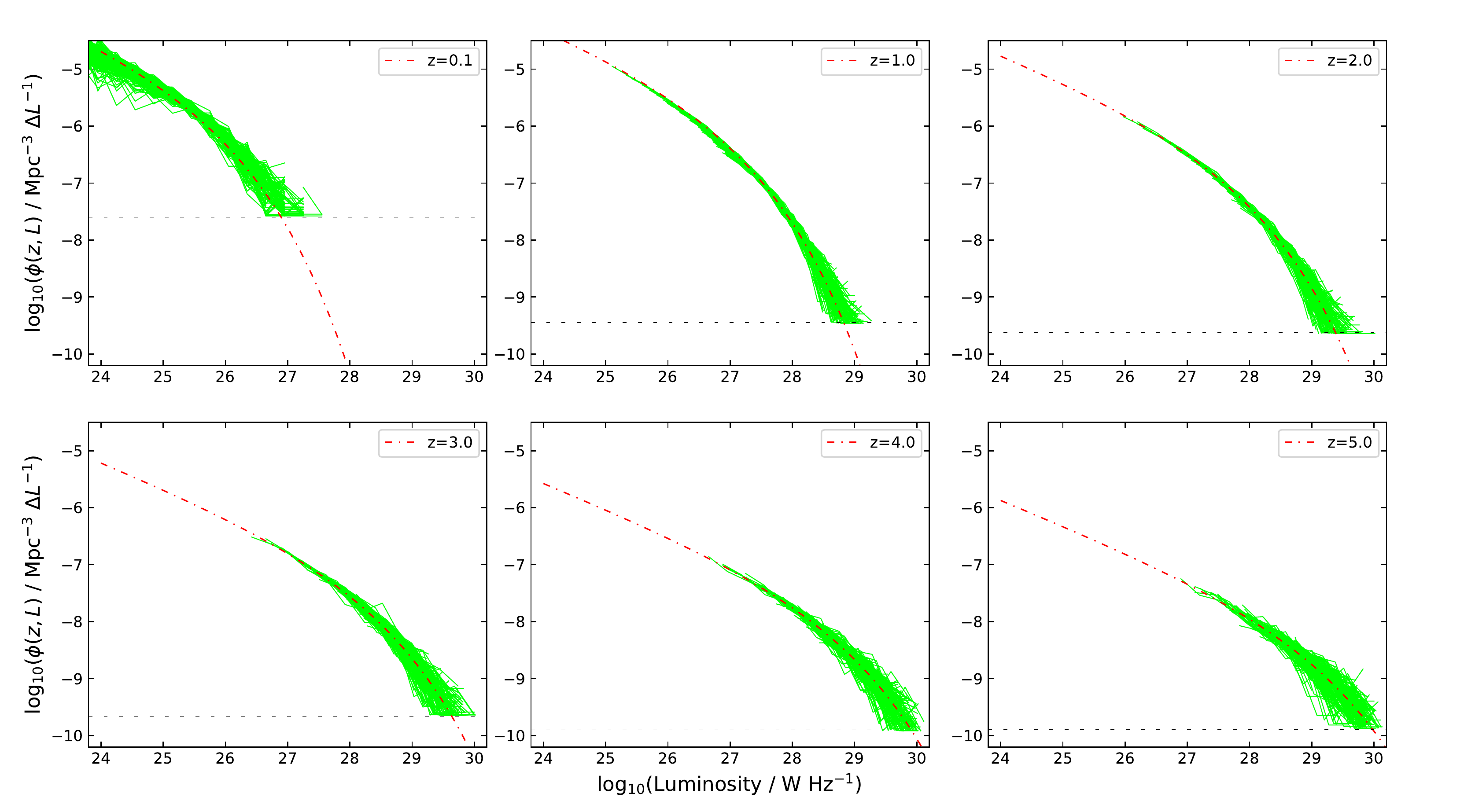}}
  \caption {\label{f8} RLFs (green lines) estimated using $\hat{\phi}_{\mathrm{bin}}$ for 200 simulated samples. The red dash-dotted lines represent the true RLF.
  The horizontal dashed lines mark the observational limits of the
  simulated surveys, below which $\hat{\phi}_{\mathrm{bin}}$ can not
  provide an extrapolated estimate.}
\end{figure*}

\section{Application to simulated data}
\label{result1}
As an illustration of the effectiveness of our KDE methods, we apply them to a simulated data set.
We use the parameterized radio luminosity function (RLF) of \citet[][model A]{2017ApJ...846...78Y} as the input LF. By choosing a flux limit
of $F_{\rm lim}=40$ mJy, and setting the solid angle $\Omega=0.456$, we ultimately simulate a flux-limited
sample of $n \thicksim 19,000$. Our sample has the redshift and luminosity limits of $(z_1=0,z_2=6)$ and $(\mathcal{L}_1=10^{22} \mathrm{W~Hz},\mathcal{L}_2=10^{30} \mathrm{W~Hz})$.
For these simulated radio sources, we assume a power-law emission spectrum of index $\alpha=0.75$
\footnote{Strictly speaking, the spectral indexes of a sample should be a distribution rather than a single value.
Nevertheless, for the steep-spectrum radio sources, the dispersion of their spectral index distribution is relatively small.
Taking an average value of $\alpha=0.75$ is a safe approximation,
where $S_{\nu} \propto \nu^{-\alpha}$.}.

\subsection{The fixed bandwidth KDE results}
Figure \ref{fig_ref} shows the RLFs for the simulated sample, estimated by the transformation KDE (denoted as $\hat{\phi}_{\mathrm{t}}$) and the
transformation-reflection KDE (denoted as $\hat{\phi}_{\mathrm{tr}}$) methods at several redshifts.
The optimal bandwidth and transformation parameters in our KDE methods are summarized in Table \ref{tab:kde}.
Their posterior probability distributions and two-dimensional (2D) confidence contours are given in Figure \ref{fig_mcmc}.
The regular unimodal feature of each distribution suggests that all the parameters are very well constrained \citep[e.g.,][]{2013ApJ...765..122Y}.

For comparison, the result measured by the traditional binned method (denoted as $\hat{\phi}_{\mathrm{bin}}$)
of \citet{2000MNRAS.311..433P} is also shown. In the literature, bins are commonly chosen somewhat arbitrarily.
\citet{2013ApSS.345..305Y} argued that this may lead to significant bias for the LF estimates close to the flux limit (see panel a of Figure \ref{bin}).
In this work we use the simple rule of thumb suggested by \citet{2013ApSS.345..305Y} to divide bins.
In the panel b of Figure \ref{bin}, we illustrate this scheme for dividing bins.

Figure \ref{fig_ref} shows that both $\hat{\phi}_{\mathrm{t}}$ and $\hat{\phi}_{\mathrm{tr}}$ are generally superior to the binned method, especially
for the high redshift ($z \geqslant 3$) LF estimation. The KDE approaches (especially $\hat{\phi}_{\mathrm{t}}$)
produce relatively smooth LFs, while the binned method produces
discontinuous estimates, being prone to have artificial bulges and
hollows.

In addition, the KDE approaches have the advantage that their measurement can be extrapolated appropriately beyond the observational limits.
In Figure \ref{fig_ref}, the vertical dashed lines mark the higher luminosity limits of the simulated survey.
Note that even beyond the limits, the extrapolated LFs can be generally acceptable approximation to the true LF.

$\hat{\phi}_{\mathrm{t}}$ and $\hat{\phi}_{\mathrm{tr}}$ have their own advantages and disadvantages. $\hat{\phi}_{\mathrm{t}}$ make smother estimates,
but it is prone to produce a slight negative bias at the faint end of the LFs. This is a typical boundary bias, suggesting that a simple transformation
method can not fully solve the boundary problem. $\hat{\phi}_{\mathrm{tr}}$ performs well at the faint end, but not ideally
at the bright end of the LFs. This is expected to be improved by using
the adaptive KDE, where the data density is used.

\subsection{The adaptive KDE results}
\label{adaptive_kde_results}

Figure \ref{fig_ada} shows the RLF and its uncertainty estimated by the transformation-reflection adaptive KDE approach (denoted as $\hat{\phi}_{\mathrm{tra}}$)
at several redshifts for the simulated sample. Given posterior probability distributions for the bandwidth and transformation parameters in our KDE methods (see Figure \ref{fig_mcmc}),
the uncertainty regions of the estimated RLF are plotted by ``fgivenx'', a public Python package of \citet{H2018}.
We find that $\hat{\phi}_{\mathrm{tra}}$ performs well for all the six redshifts, and it achieves an excellent approximation to the true LF.
Comparing with Figure \ref{fig_ref}, we notice that $\hat{\phi}_{\mathrm{tra}}$ can overcome the shortcomings of $\hat{\phi}_{\mathrm{tr}}$ and $\hat{\phi}_{\mathrm{tr}}$.

In order to rule out the possibility that our adaptive KDE estimator only gives the good result by chance, we simulate 200 flux-limited samples with
their flux limits randomly drawn between $10^{-2.5}$ and $10^{-0.5}$ Jy. All the simulated samples share the same input LF
\citep[the model A RLF of][]{2017ApJ...846...78Y}.
By adjusting the simulated solid angle,
we can control the size ($n$) of each sample. We let the sample size be linearly proportional to the logarithm
of its flux limit. Finally, for the 200 simulated samples, their sizes
are randomly distributed between 2,000 and 40,000 sources.

Figures \ref{f7} and \ref{f8} shows the RLFs estimated by $\hat{\phi}_{\mathrm{tra}}$ and $\hat{\phi}_{\mathrm{bin}}$, respectively, based on 200 simulated samples.
Obviously, the $\hat{\phi}_{\mathrm{tra}}$ estimator shows better stability and reliability in performance than the binned method.
In most cases, we find that the LFs estimated by $\hat{\phi}_{\mathrm{bin}}$ look like irregular sawtooth, randomly leaping up and sloping down.
This may mislead the observer to use wrong parametric form to model the LF. Our $\hat{\phi}_{\mathrm{tra}}$ estimator does not have such drawback.
In fact, this is why the KDE method is very popular in modern statistics: it can produce smoother estimation which converge to the true density faster \citep{Wasserman2006}.

We note that in Figure \ref{f7} the adaptive KDE method is slightly asymmetrically biased to values above the true LF compared to below the true LF
(also see Figure \ref{fig_ada}), especially for the tails where LFs are extrapolated. This is because the functions of adaptive kernels given in Equation (\ref{hx}) only have one adjustable parameter, $\beta$, and cannot let the estimated LF go too steep too soon in the extremely low number density regime. The above effect does not matter as it is only visible for the extrapolated LFs.

\begin{figure}[!h]
\centering
\includegraphics[width=1.05\columnwidth]{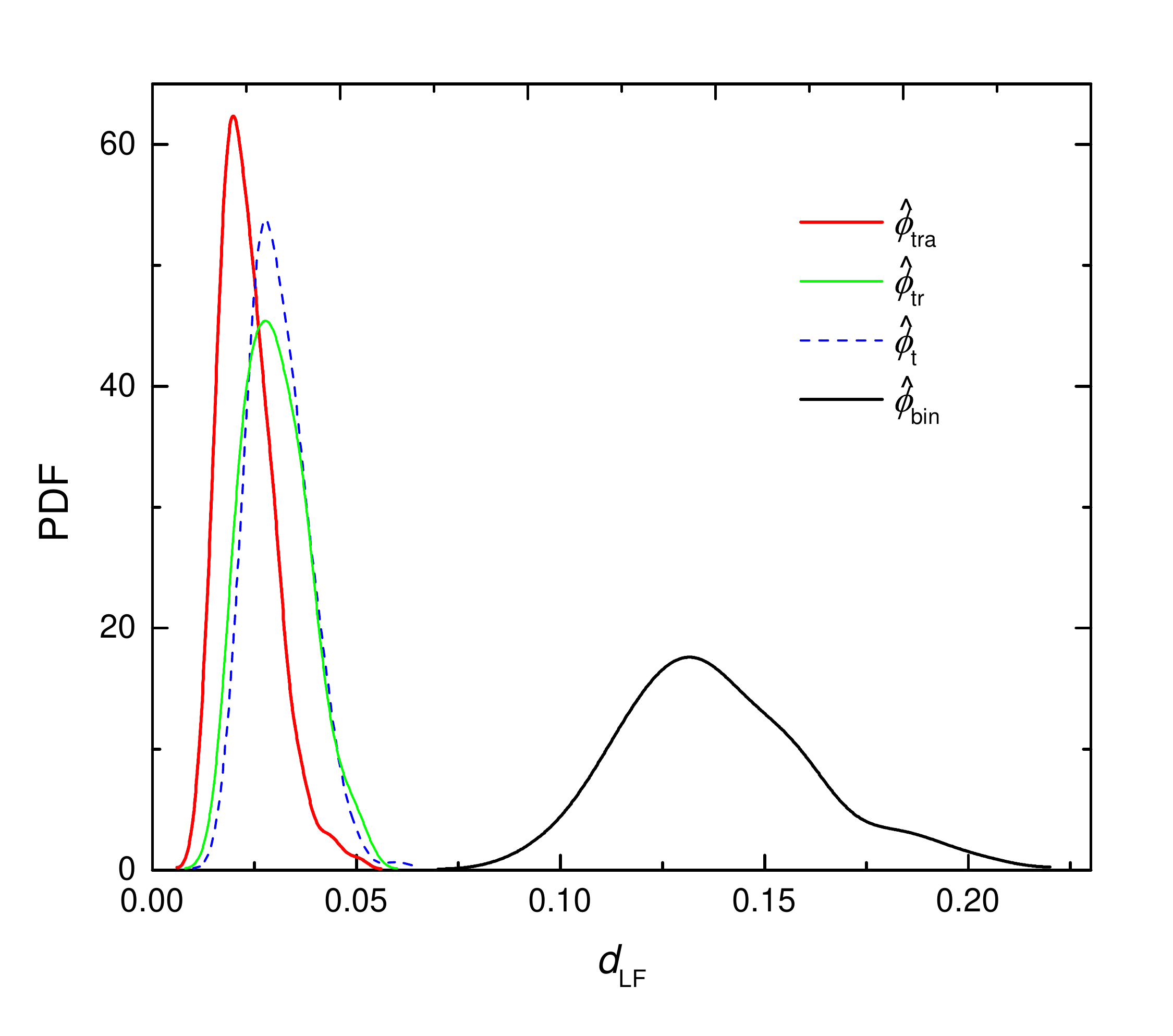}
\caption{Distributions of $d_{\mathrm{LF}}$ for different LF
  estimators for the 200 simulated samples. The red solid, green solid, blue dashed, and black solid curves correspond to the
$\hat{\phi}_{\mathrm{tra}}$, $\hat{\phi}_{\mathrm{tr}}$, $\hat{\phi}_{\mathrm{t}}$, and $\hat{\phi}_{\mathrm{bin}}$ estimators, respectively.}
\label{pdf_dLF}
\end{figure}

\begin{figure}[!h]
\centering
\includegraphics[width=1.05\columnwidth]{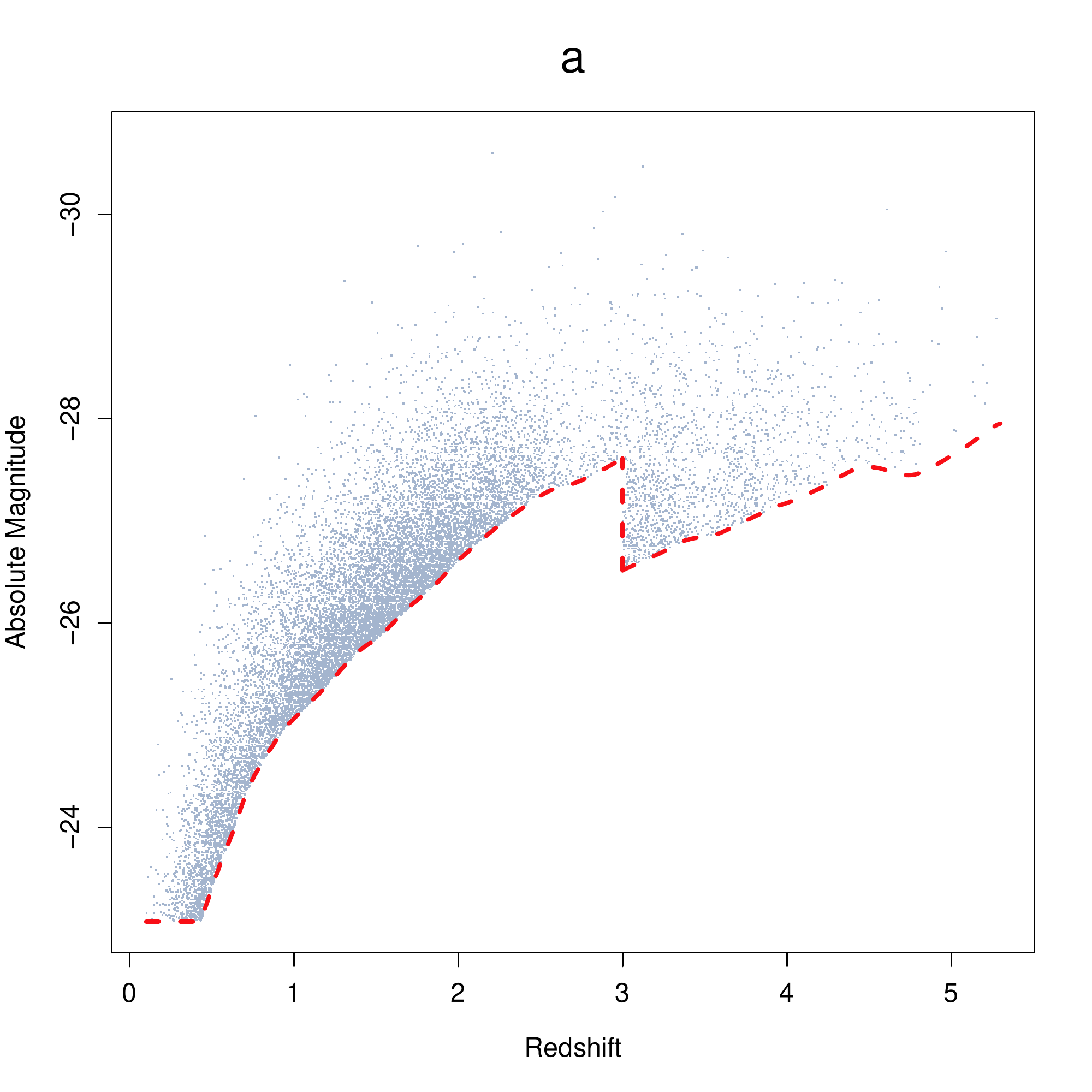}
\includegraphics[width=1.05\columnwidth]{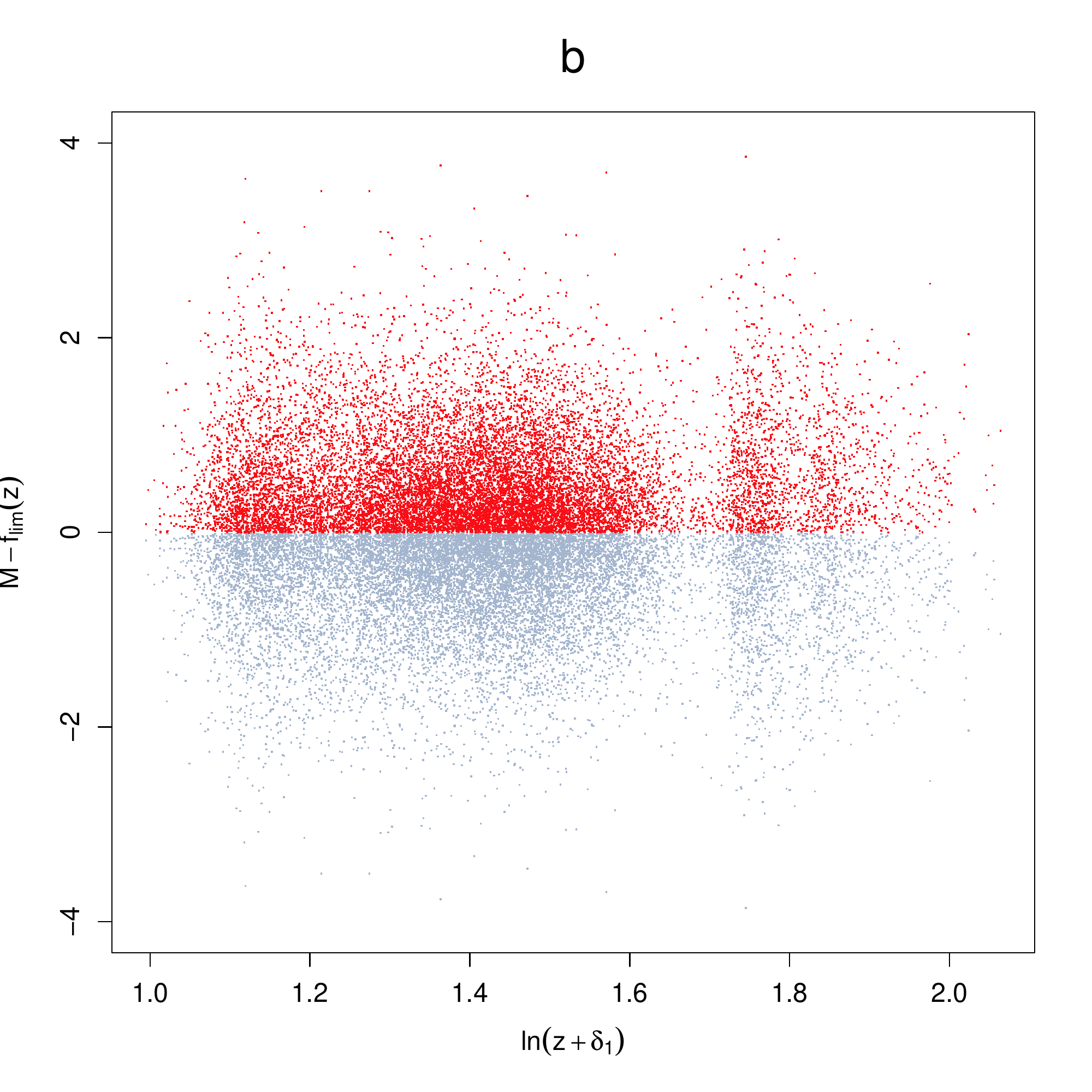}
\caption{Panel a: The quasar sample used in this analysis. The data are from \citet{2006AJ....131.2766R}. The red dashed line shows the truncation boundary, defined by $f_{\mathrm{lim}}(z)$.
Panel b shows how the quasar data look like after transformation (light grey points) by Equation (\ref{trans_M}), and adding the reflection points (red points).}
\label{fig_SDSS_quasar}
\end{figure}

\begin{figure*}
\centerline{
\includegraphics[scale=0.75,angle=0]{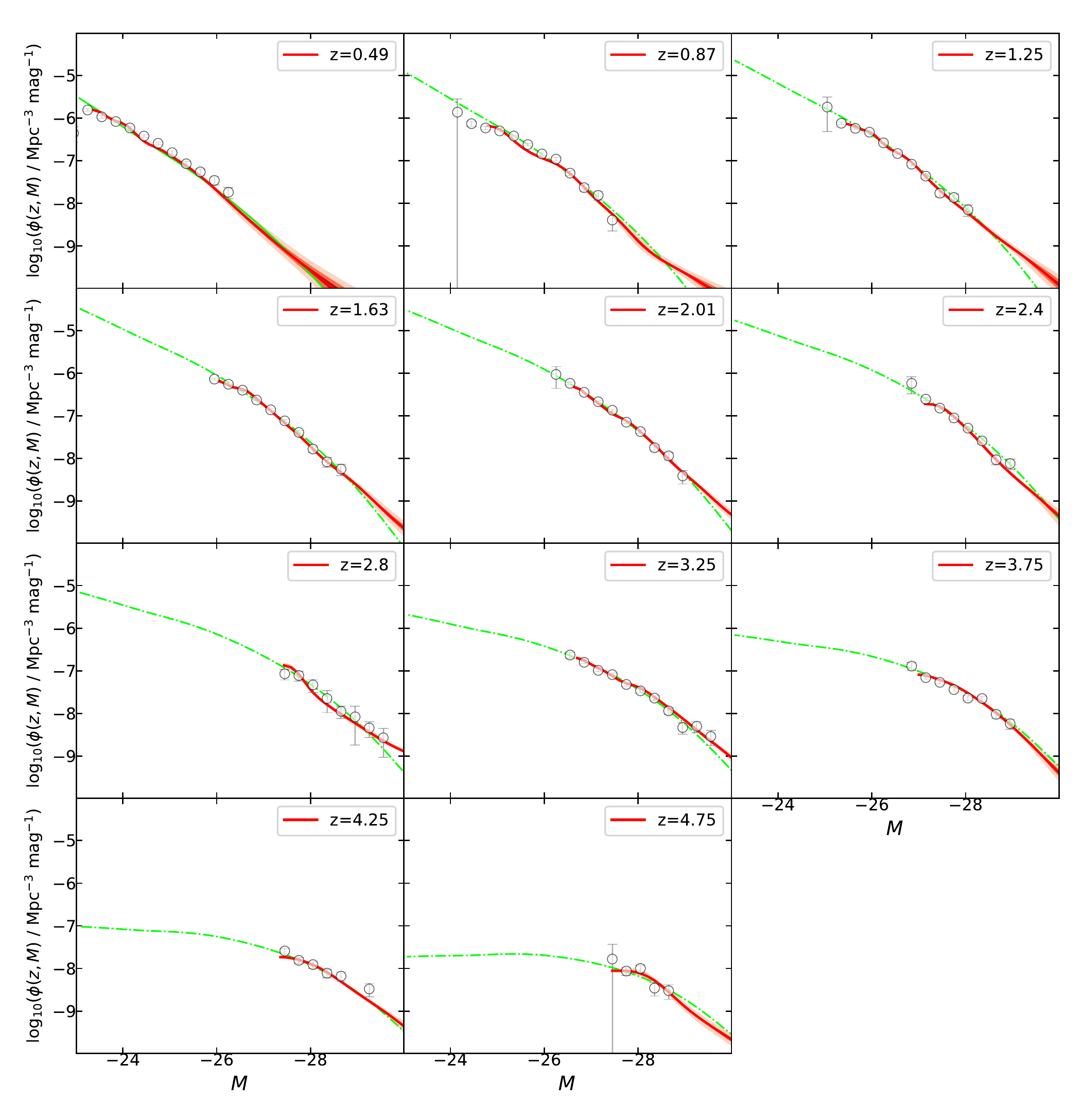}}
\caption {\label{fig_quasar_LF} Quasar LF estimated by our transformation-reflection adaptive KDE approach (red solid lines) at different redshifts. The light shaded areas take into account the 3 $\sigma$ error bands. Comparisons are made with the semi-parametric estimates (green dash dotted lines) given in \citet{2007ApJ...661..703S},
and the binned estimates (light gray circles with error bars) given in \citet{2006AJ....131.2766R}.}
\end{figure*}

\begin{table}
\renewcommand{\arraystretch}{1.3}
\caption{Parameters of the KDE for the simulated sample.}
\begin{center}
\begin{tabular}{cc}
\hline\hline
Method    & Parameters\\
\hline
$\hat{\phi}_{\mathrm{t}}$   & $\ln\delta_1=-1.262_{-0.322}^{+0.374}$;\,$\ln\delta_2=-3.064_{-0.200}^{+0.190}$ \\
                            & \underline{$h_1=0.123_{-0.014}^{+0.011}$;\,$h_2=0.127_{-0.008}^{+0.009}$}\\

$\hat{\phi}_{\mathrm{tr}}$  & $\ln\delta_1=-0.892_{-0.402}^{+0.207}$;\, \\
                            & \underline{$h_1=0.100_{-0.009}^{+0.014}$;\,$h_2=0.105_{-0.005}^{+0.004}$}\\

$\hat{\phi}_{\mathrm{tra}}$ & $\ln\tilde{\delta}_1=-0.892$;\,$\tilde{h}_1=0.100$;\,$\tilde{h}_2=0.105$ \\
                            & $h_{10}=0.080_{-0.003}^{+0.007}$; \,$h_{20}= 0.070_{-0.001}^{+0.003}$;  \\
                            & $\beta=0.237_{-0.019}^{+0.018}$ \\
\hline
\end{tabular}
\end{center}
~~$\mathbf{Notes}$. $\hat{\phi}_{\mathrm{tra}}$ is implemented on the basis of $\hat{\phi}_{\mathrm{tr}}$, and $\tilde{\delta}_1$, $\tilde{h}_1$ \& $\tilde{h}_2$
inherit the parameter values of $\hat{\phi}_{\mathrm{tr}}$. Parameter errors correspond to the $68\%$ confidence level.
Parameters without an error estimate were kept fixed during the fitting stage.
\label{tab:kde}
\end{table}

\begin{table}[!t]
\tablewidth{0pt}
\renewcommand{\arraystretch}{1.5}
\caption{The statistic $d_{\mathrm{LF}}$}
\begin{center}
\begin{tabular}{ccccc}
\hline\hline

\colhead{} & \colhead{$\hat{\phi}_{\mathrm{bin}}$} & \colhead{$\hat{\phi}_{\mathrm{t}}$}  & \colhead{$\hat{\phi}_{\mathrm{tr}}$} & \colhead{$\hat{\phi}_{\mathrm{tra}}$} \\
\hline
 $d_{\mathrm{LF}}$                &0.0944  & 0.0239  & 0.0193  & 0.0157 \\
 $\langle d_{\mathrm{LF}}\rangle$ &0.1387  & 0.0311  & 0.0305  & 0.0227 \\
\hline
\end{tabular}
\end{center}
~~$\mathbf{Notes}$. $d_{\mathrm{LF}}$ is calculated for the simulated sample described in section \ref{result1}. $\langle d_{\mathrm{LF}}\rangle$ is the mean value for 200 simulated samples.
\label{tab:dLF}
\end{table}

\begin{figure*}
\includegraphics[height=8cm,width=8.25cm]{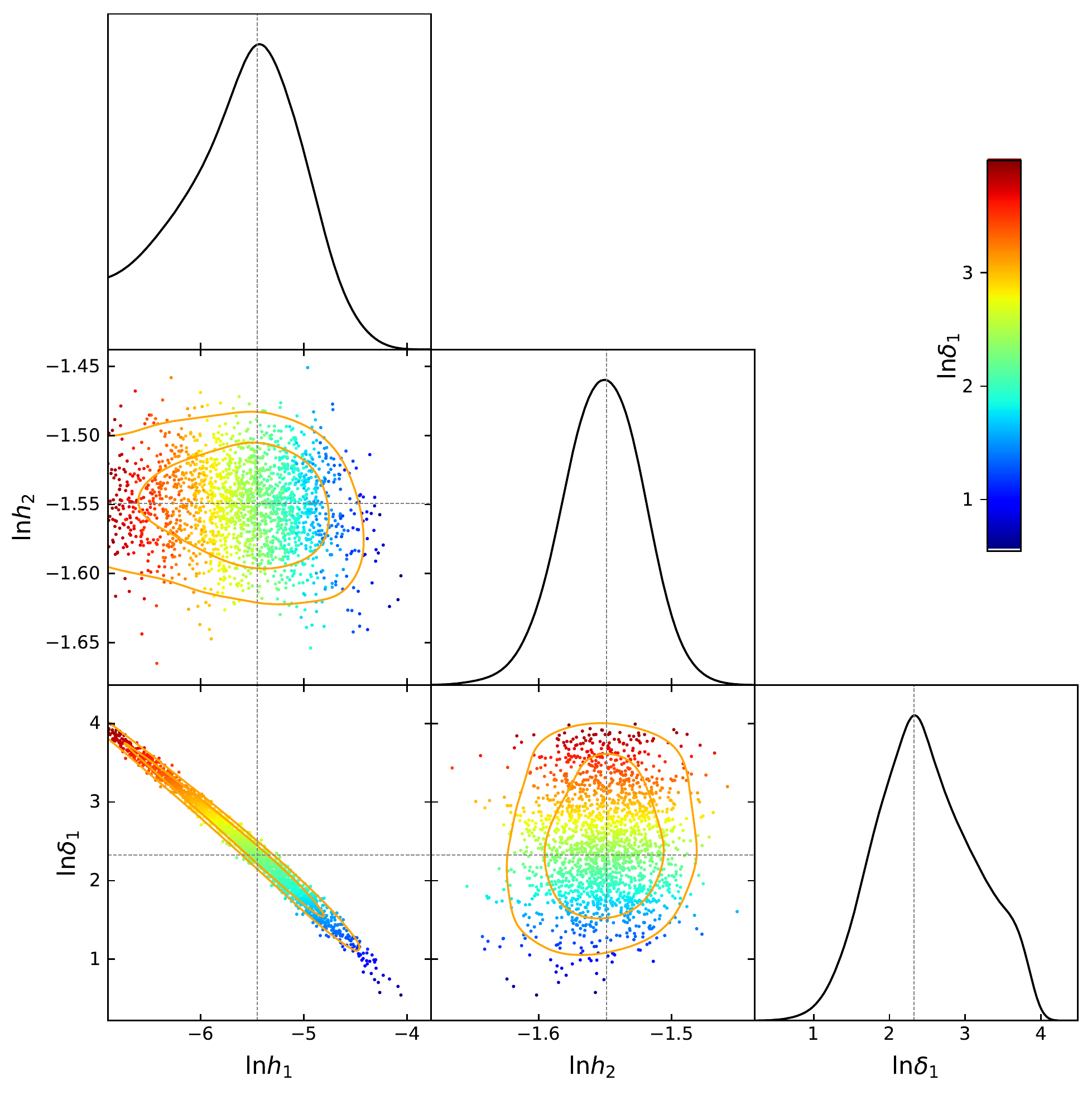}  \qquad
\includegraphics[height=8cm,width=8.25cm]{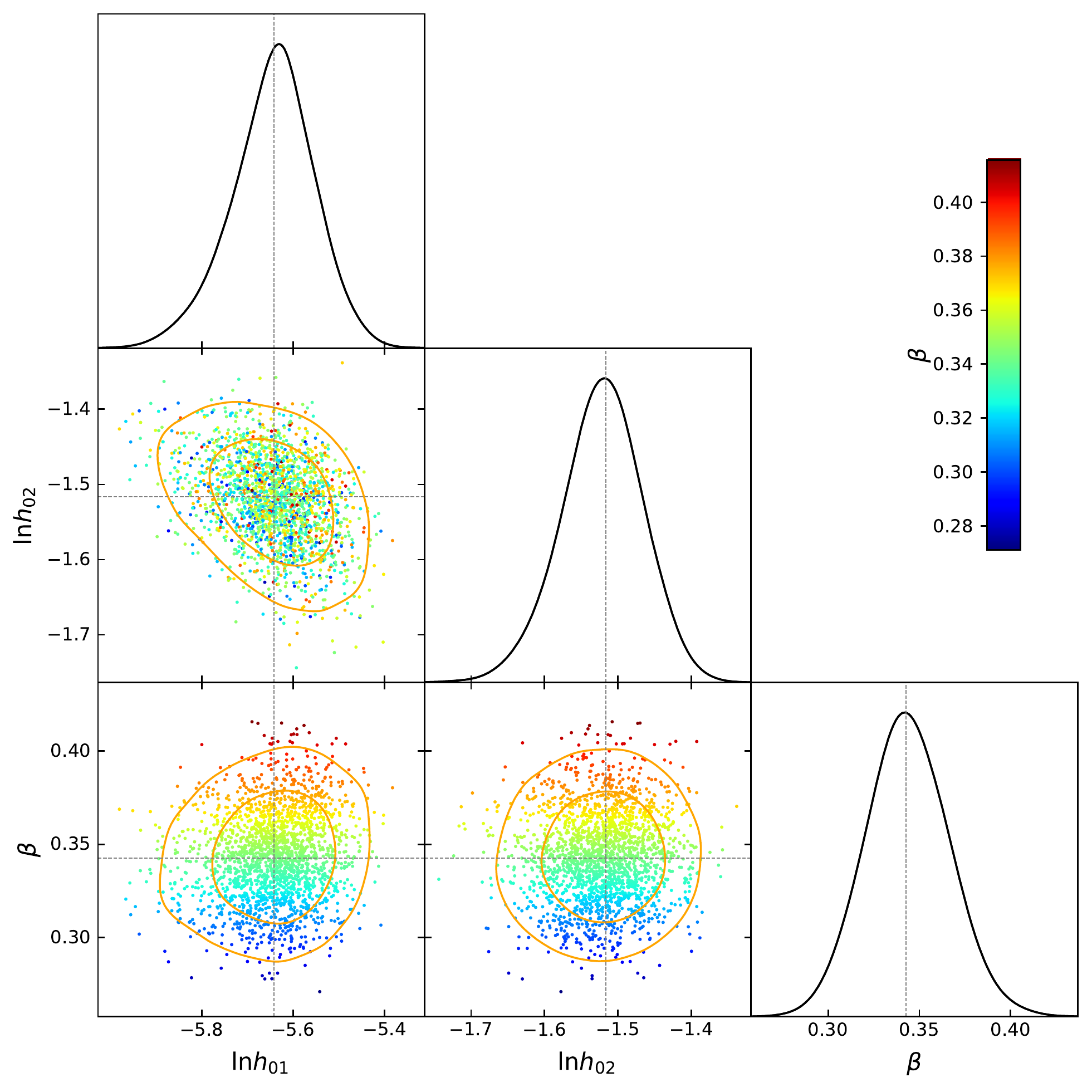}
\caption{Marginalized 1D and 2D posterior distributions of parameters for estimating the quasar LF. The contours containing 68\% and 95\% of the probability.
The meaning of coloured points and vertical lines are the same with that in Figure \ref{fig_mcmc}. The left and right panels correspond to
$\hat{\phi}_{\mathrm{tr}}$ and $\hat{\phi}_{\mathrm{tra}}$ respectively.}
\label{fig_mcmc_SDSS}
\end{figure*}

\subsection{Quantitative evaluation of performance}
\label{qep}
To quantify the performance of each LF estimator, we define a statistic $d_{\mathrm{LF}}$ which measures the discrepancy of the
estimated LF $\hat{\phi}$ from the true LF $\phi$. If a large number of random vectors, denoted by
\{$(z_i,L_i)$\}$_{i=1}^{n}$, can be drawn from $\phi(z,L)$, $d_{\mathrm{LF}}$ can be estimated \citep[e.g.,][]{Zhang2006} by
\begin{eqnarray}
\label{dkl1}
d_{\mathrm{LF}}(\phi,\hat{\phi})=\frac{1}{n}\sum_{i=1}^{n}\left| \log_{10}\left(\phi(z_i,L_i)/\hat{\phi}(z_i,L_i)\right) \right|.
\end{eqnarray}
For the $\hat{\phi}_{\mathrm{bin}}$ estimator, estimates are given only at some discontinuous points, i.e., the centers of each bin.
The statistic $d_{\mathrm{LF}}$ is estimated by
\begin{eqnarray}
\label{dkl2}
d_{\mathrm{LF}}(\phi,\hat{\phi}_{\mathrm{bin}})=\frac{1}{N_{\mathrm{bin}}}\sum_{j=1}^{N_{\mathrm{bin}}}
\left |  \log_{10} \left( \frac{\phi(z_{j}^c,L_{j}^c)}{\hat{\phi}_{\mathrm{bin}}(z_{j}^c,L_{j}^c)}  \right) \right | ,
\end{eqnarray}
where $N_{\mathrm{bin}}$ is the number of bins, and $(z_{j}^c,L_{j}^c)$ locates the center of $j$th bin.
The edges of the redshift bins are $0.0, 0.2, 0.5, 0.8, 1.2, 1.8, 2.2, 2.7, 3.3, 3.8, 4.2, 4.7, 5.3, 6.0$. The logarithmic luminosity bins
are in increments of 0.3.

In Table \ref{tab:dLF}, we show the statistic $d_{\mathrm{LF}}$ of different LF estimators calculated for our simulated sample.
Also the mean values $\langle d_{\mathrm{LF}}\rangle$ for 200 simulated samples are given. The statistic $d_{\mathrm{LF}}$ of all
our KDE estimators, especially $\hat{\phi}_{\mathrm{tra}}$, are significantly better than that of $\hat{\phi}_{\mathrm{bin}}$.
$d_{\mathrm{LF}}$ can be understood as the typical error of a LF estimator. In this sense, the $\hat{\phi}_{\mathrm{tra}}$ estimator
improves the accuracy by nearly an order of magnitude compared to $\hat{\phi}_{\mathrm{bin}}$.
Figure \ref{pdf_dLF} shows the distributions of $d_{\mathrm{LF}}$ of different LF estimators for 200 simulated samples.
The smaller dispersion of distributions for our KDE methods suggests that they all have significantly better stability
than the $\hat{\phi}_{\mathrm{bin}}$ estimator, as expected.

\section{Application to SDSS Quasar Sample}
\label{result2}
In this section, we apply our adaptive KDE method to the quasar sample of \citet{2006AJ....131.2766R}. This sample consists of 15,343 quasars
within an effective area of 1622 deg$^2$ that was a subset of the broad-line quasar sample from Sloan Digital Sky Survey (SDSS) Data Release 3.
Following \citet{2007ApJ...661..703S}, we remove 62 quasars with absolute magnitude (denoted as $M$) $> -23.075$, and 224 additional quasars that
fall in an extremely poorly sampled region. After this truncation, there are 15,057 quasars remaining \citep[see the section 2 of][]{2007ApJ...661..703S}.
In Figure \ref{fig_SDSS_quasar}a, we show these quasars (light grey points) in the $M-z$ plane as well as the truncation boundary (red dashed line).
According to \citet{2006AJ....131.2766R}, the sample was not assumed to be complete within the truncated region. Thus, each object was assigned a weight
($w_i$, inverse of the value of the selection function) depending on its redshift and apparent magnitude. The selection function did not have an analytical form
and was approximated via simulations \citep[see][for
details]{2006AJ....131.2766R}.

The quasar LF, $\phi(z,M)$, is defined by simply replacing $L$ with $M$ in Equation (\ref{LFdf}). To estimate the quasar LF using the transformation-reflection
adaptive KDE approach, we transform the original quasar data by
\begin{eqnarray}
\label{trans_M}
x=\ln(z+\delta_1), ~\mathrm{and}, ~ y=M-f_{\mathrm{lim}}(z),
\end{eqnarray}
where $f_{\mathrm{lim}}(z)$ is the function defining the truncation boundary of the quasar sample. It does not have an analytical form.
\citet{2007ApJ...661..703S} provided the values of $f_{\mathrm{lim}}(z)$ at a series of discrete points, based on which, we can calculate
$f_{\mathrm{lim}}(z)$ by a linear interpolation. Figure \ref{fig_SDSS_quasar}b shows how the quasar data look after transformation (light grey points)
and adding the reflection points (red points). There is a clear dip in the density of quasars at $z\thickapprox 2.7$, which is especially obvious after the data
are transformed. This is not surprising, since the selection function is particularly low at redshift 2.7, where quasars have colors very similar
to A¨CF stars \citep{2006AJ....131.2766R}. Hence, the weighting due to the selection function needs to be taken
into account in using our adaptive KDE approach. The detailed formulas for considering the weighting can be found in Appendix \ref{A2}.

Figure \ref{fig_quasar_LF} shows the quasar LF estimated by our transformation-reflection adaptive KDE approach (red solid lines) at several redshifts.
The optimal bandwidth and transformation parameters of the KDE are summarized in Table \ref{tab:quasar_LF}.
Their posterior probability distributions and two-dimensional (2D) confidence contours are given in Figure \ref{fig_mcmc_SDSS}.
Comparisons are made with the semi-parametric estimates (green lines with error bars) given in \citet{2007ApJ...661..703S},
and the binned estimates (light gray circles with error bars) given in \citet{2006AJ....131.2766R}.
Our result is in good agreement with the two previous
estimates. \citet{2007ApJ...661..703S} gives estimates beyond the lower luminosity limit, which can be ascribed to the assumed parametric form $\mathbf{h}(z,M,\theta)$ used in their method. Both the \citet{2007ApJ...661..703S} method and
our adaptive KDE estimator can give extrapolations on the LF beyond the higher luminosity limits, and the two results are generally consistent with each other.
The ability of extrapolating LFs beyond currently observable luminosities and redshifts is very useful,
as this may guide the design of future surveys \citep{2018ApJ...869...96C}. Overall, our method achieves estimates comparable to that of \citet{2007ApJ...661..703S},
but makes no assumptions about the parametric form of the LF.

\begin{table}
\renewcommand{\arraystretch}{1.3}
\caption{Parameters of the KDE for the quasar sample}
\begin{center}
\begin{tabular}{cc}
\hline\hline
Method    & Parameters\\
\hline
$\hat{\phi}_{\mathrm{tr}}$  & $\ln\delta_1=2.275_{-0.443}^{+0.947}$;\, \\
                            & \underline{$\ln h_1=-5.408_{-0.837}^{+0.364}$;\,$\ln h_2=-1.551_{-0.031}^{+0.029}$}\\

$\hat{\phi}_{\mathrm{tra}}$ & $\ln\tilde{\delta}_1=2.275$;\,$\ln\tilde{h}_1=-5.408$;\,$\ln\tilde{h}_2=-1.551$ \\
                            & $\ln h_{10}=-5.629_{-0.113}^{+0.073}$; \,$\ln h_{20}=-1.517_{-0.060}^{+0.053}$;\, \\
                            & $\beta=0.343_{-0.023}^{+0.023}$ \\
\hline
\end{tabular}
\end{center}
$\mathbf{Notes}$. Parameter errors correspond to the $68\%$ confidence level.
Parameters without an error estimate were kept fixed during the fitting stage.
\label{tab:quasar_LF}
\end{table}

\section{Discussion}
\label{discussion}
\subsection{Comparison with Previous Methods}
Hitherto, the classical binned estimator ($\hat{\phi}_{\mathrm{bin}}$) is still the most popular non-parametric method.
The original version of $\hat{\phi}_{\mathrm{bin}}$ is the famous $1/V_{\mathrm{max}}$ estimator \citep{1968ApJ...151..393S}.
It was originally a variation of the $V/V_{\mathrm{max}}$ test \citep{1968ApJ...151..393S},
while $V/V_{\mathrm{max}}$ is essentially a completeness estimator \citep[see][]{2011A&ARv..19...41J}.
This makes it impossible for the $1/V_{\mathrm{max}}$ method to be a mathematically rigorous density estimator.
The $1/V_{\mathrm{max}}$ estimator can not accurately calculate the surveyed regions
for objects close to the flux limit of their parent sample and thus produces a significant systematic error \citep[see][]{2000MNRAS.311..433P,2013ApSS.345..305Y}.
To improve the $1/V_{\mathrm{max}}$ estimator, \citet{2000MNRAS.311..433P} proposed a new method that rests on a stronger mathematical foundation.
The new method is superior to the $1/V_{\mathrm{max}}$ estimator \citep[e.g.,][]{2013ApSS.345..305Y}.
For the LF at the center of a bin with a luminosity interval ($L_{\mathrm{min}}$,$L_{\mathrm{max}}$) and a redshift interval ($z_{\mathrm{min}}$, $z_{\mathrm{max}}$),
the new estimator gave
\begin{eqnarray}
\label{phi_pc}
\phi_{\mathrm{est}}=\frac{N}{\int_{L_{\mathrm{min}}}^{L_{\mathrm{max}}}\int_{z_{1}}^{z_{\mathrm{max}}(L)}\frac{dV}{dz}dzdL}
\end{eqnarray}
where $N$ is the number of sources detected within the bin. The double integral in Equation (\ref{phi_pc}) naturally considers the truncation boundary as discussed in
section \ref{truncation_boundary}. In this text, we do not strictly distinguish among the \citet{2000MNRAS.311..433P} method, $1/V_{\mathrm{max}}$ and $1/V_{\mathrm{a}}$
\citep[the generalized version of $1/V_{\mathrm{max}}$ used for multiple samples,][]{1980ApJ...235..694A}.
They are collectively referred to as $\hat{\phi}_{\mathrm{bin}}$.

$\hat{\phi}_{\mathrm{bin}}$ is widely acknowledged for its simplicity and ease of implementation.
However, it has some sever drawbacks. First, its estimate depends on the dividing of bins but currently there are no effective rules
to guide the binning. Second, it produces discontinuous estimates, and the discontinuities of the estimate are not due to the underlying LF,
but are only an artifact of the chosen bin locations. These discontinuities make it very difficult to grasp the structure of the data.
Third, the binning of data undoubtedly leads to information loss and potential biases can be caused by evolution within the bins.
Fourth, $\hat{\phi}_{\mathrm{bin}}$ is a bivariate estimator and it can not be properly extended to the trivariate situation.
Since the number of bins grows exponentially with the number of dimensions, in higher dimensions one would require many more
data points or else most of the bins would be empty \citep[the so-called curse of dimensionality,][]{Gramacki2018}.
A trivariate LF estimator is necessary in the situation when
additional quantities \citep[such as photon index, ][]{2012ApJ...751..108A} besides
the redshift and luminosity are incorporated into the LF analysis to tackle complex $K$-corrections \citep[see][]{2016ApJ...829...95Y}.
In this case, there is risk of using a low dimensional tool to deal with higher dimensional problems.
These drawbacks prevent $\hat{\phi}_{\mathrm{bin}}$ from being a precise estimator.
Usually, the estimates of $\hat{\phi}_{\mathrm{bin}}$ are used to provide guidance or calibration for modeling the LF.
If the $\hat{\phi}_{\mathrm{bin}}$ result itself is not accurate, the parametric description can not be reliable.

From a mathematical perspective, the $\hat{\phi}_{\mathrm{bin}}$ estimator is a kind of two-dimensional histogram.
All the drawbacks of $\hat{\phi}_{\mathrm{bin}}$ are inherited from the histogram.
In the mathematical community, the histogram has been a outdated density estimator except for rapid visualization of results in one or two
dimensions \citep[e.g,][]{Gramacki2018}. To conquer the shortcomings of histograms, mathematicians have developed some other non-parametric
estimators, such as smoothing histograms, Parzen windows, k-nearest neighbors, KDE, etc. Among of these, KDE is the most popular density
estimator. This is why we develop the new LF estimator within a KDE framework. Our KDE methods can overcome all the drawbacks of $\hat{\phi}_{\mathrm{bin}}$
and significantly improve the accuracy.

The \citet{1971MNRAS.155...95L} $C^{-}$ method and its variants \citep[e.g.,][]{1992ApJ...399..345E,1993ApJ...416..450C}
is another important estimator. It overcomes some of the shortcomings in $\hat{\phi}_{\mathrm{bin}}$. Recently, \citet{2014ApJ...786..109S}
extended $C^{-}$ to the trivariate scenario, and presented the redshift evolutions and distributions of the gamma-ray luminosity
and photon spectral index of flat spectrum radio quasars. This suggested that $C^{-}$ has a good extensibility in dimension.
However, the direct product of $C^{-}$ is the cumulative distribution function of the LF but not the LF itself,
and this is inconvenient. A more serious disadvantage is that $C^{-}$ typically assume that luminosity and redshift are statistically independent.
In the actual application, one inevitably has to introduce a
parametrised model that removes the correlation of luminosities and redshifts \citep[e.g.,][]{2014ApJ...786..109S}.

Motivated by the potential hazards and pitfalls existing in the above traditional estimators, approaches based on more innovative statistics have been developed.
One of the typical representative is the semi-parametric approach of \citet{2007ApJ...661..703S}. Schafer decomposed the brivariate density $\phi(z,M)$ into
\begin{eqnarray}
\label{schafer}
\log \phi(z,M,\theta)=f(z) + g(M) + h(z,M,\theta),
\end{eqnarray}
where $f(z)$ and $g(z)$ are determined non-parametrically, while $h(z,M,\theta)$ has a parametric form.
From Figure \ref{fig_quasar_LF}, we find that our adaptive KDE method
achieves estimates comparable to that of \citet{2007ApJ...661..703S}
while making far fewer assumptions about the shape of LF.
The Bayesian approach of \citet{2008ApJ...682..874K} is another estimator based on innovative statistics.
This method is similar to ours in some aspects, e.g., both are within a Bayesian framework and using the MCMC algorithms for estimating the parameters;
both modeling the LF as a mixture of Gaussian functions. The difference is that we arrange a relatively unified gaussian function at each data point and
the final LF is the sum of all the gaussian functions, while
\citet{2008ApJ...682..874K} use far fewer (typically~$\thicksim \!\!3-6$) gaussian functions and each of them has
six free parameters for adjusting the shape and location. Therefore, our method has much fewer (typically~$\thicksim \!\!3-6$) parameters than theirs (typically$\thicksim22-40$).
In addition, their method requires much more critical prior information for parameters to aid the convergence of MCMC. Generally they assume that the LF is unimodal, thus
their prior distributions are constructed to place more probability on situations where the Gaussian functions are close together \citep[see the Fig. 3 of][]{2008ApJ...682..874K}.
Our KDE method does not impose any preliminary assumptions on the shape of LFs, and simply use uniform (so-called ``uninformative'') priors for the parameters to run the MCMC.
Therefore, our method is more flexible.

\subsection{Extensibility}
\label{tr_3d}
Our KDE methods can be easily extended to the trivariate LF
situation. Below we give an example for employing the
$\hat{\phi}_{\mathrm{tr}}$ estimator in this case.
Suppose we observe $n$ objects $\{(\alpha_i,z_i,L_i), i=1,2,...,n\}$ within a survey region $W$, and
$(\boldmath{\alpha},\boldmath{z}, \boldmath{L}) \in W \subset \mathbf{R}^3$.
$\alpha_i,z_i,L_i$ are the single power-law spectral index, redshift, and luminosity of the $i$th object.
Transform the observed data by
\begin{eqnarray}
\label{trans_3d}
\alpha=\alpha, ~x=\ln(z+\delta_1), ~\mathrm{and},~ y=L-f_{\mathrm{lim}}(\alpha,z),
\end{eqnarray}
where $f_{\mathrm{lim}}(\alpha,z)$ defines the truncation boundary of the sample, and is given by
\begin{eqnarray}
\label{f_lim_alpha}
f_{\mathrm{lim}}(\alpha,z)=4\pi d_L^{2}(z)(1/K(\alpha,z))F_{\mathrm{lim}},
\end{eqnarray}
where $d_L(z)$ is the luminosity distance, $F_{\mathrm{lim}}$ is the
survey flux limit, and $K(\alpha,z)$ represents the $K$-correction,
where $K(\alpha,z)=(1+z)^{1-\alpha}$ for a power-law emission spectrum of index $\alpha$.
If the emission spectrum is not a power-law, $K(\alpha,z)$ would have
a different functional form.
Thus we reserve the possibility that $\alpha$ can also represent other quantities to determine the $K$-correction.
In Appendix \ref{trivariate_KDE}, we give the procedure of use the transformation-reflection KDE for trivariate LFs.
The adaptive KDE formulas can be easily derived.

\subsection{Future development}
In mathematics, KDE is a very effective smoothing technique with many practical applications,
and analysis of the KDE methods is an ongoing research task.
New research on bandwidth selection, reducing the boundary bias, multivariate KDE, and fast KDE algorithm for big data, etc.,
are continuously emerging \citep[e.g.,][]{Cheng2019,Igarashia2019}. The fast KDE algorithm is especially relevant to astronomical surveys in the near future.
From Equation (\ref{kde1}), we note that the computational complexity of KDE is of $O(n^2)$, and it will be
computationally expensive for large datasets and higher dimensions. In recent years, techniques such as using the fast Fourier transform (FFT)
have been proposed to accelerate the KDE computations \citep[e.g.,][]{Gramacki2017a,Gramacki2017b,DB2018}.
These ongoing developments in the KDE theory means that our LF
estimator could be seen as in starting method with space for upgrading, and enable it to flexibly deal with various LF estimating problems in the future surveys.

\section{Summary}
\label{sum}
We summarize the important points of this work as follows.
\begin{enumerate}
  \item We propose a flexible method of estimating luminosity functions (LFs) based on the kernel density estimation (KDE), the most popular
  non-parametric approach to density estimation developed in modern statistics.
  In view that the bandwidth selection is crucial for the KDE, we develop a new likelihood cross-validation
  criterion for selecting optimal bandwidth, based on the well known likelihood function of \citet{1983ApJ...269...35M}.

  \item One challenge in applying the KDE to LF estimation is how to treat the boundary bias problem, since astronomical surveys usually
  obtain truncated sample of objects due to the observational limitations. We use two solutions, the transformation KDE method ($\hat{\phi}_{\mathrm{t}}$),
  and the transformation-reflection KDE method ($\hat{\phi}_{\mathrm{tr}}$) to reduce the boundary bias.
  The posterior probability distribution of bandwidth and transformation parameters for $\hat{\phi}_{\mathrm{t}}$ and $\hat{\phi}_{\mathrm{t}}$
  are derived within a Markov chain Monte Carlo (MCMC) sampling procedure.

  \item Based on a Monte Carlo simulation, we find that both the performance of $\hat{\phi}_{\mathrm{t}}$ and $\hat{\phi}_{\mathrm{tr}}$
  are superior to the traditional binned method.

  \item To further improve the performance of our KDE methods, we develop the transformation-reflection adaptive KDE method ($\hat{\phi}_{\mathrm{tra}}$).
  Monte Carlo simulations show that it achieves an excellent
  approximation to the true LF, with a good stability and reliability
  in performance, with accuracy of
  around an order of magnitude better than for the binned method.

  \item We apply our adaptive KDE method to a quasar sample and obtain consistent results to
   the rigorous determination of \citet{2007ApJ...661..703S}, while
   making far fewer assumptions about the shape of LF.

  \item The KDE method we develop has the advantages of both parametric and non-parametric methods. It (1) does not assume a particular parametric form for the LF;
(2) does not require dividing the data into arbitrary bins, thereby reducing information loss and preventing potential biases caused by evolution within the bins;
(3) produces smooth and continuous estimates; (4) utilizes the Bayesian method to maximize the exploitation of data information;
(5) is a new development but with opportunities for upgrading, making it have the flexibility to deal with various LF estimating problems in the future surveys.

\end{enumerate}

\acknowledgments
We thank the anonymous reviewer for the many comments
and suggestions leading to a clearer description of these results.
We acknowledge the financial support from the National Natural Science
Foundation of China (grant Nos. 11603066, U1738124, 11573060 and 11661161010),
and Yunnan Natural Science Foundation (Nos. 2019FB008 and 2019FB009).
We would like to thank Xibin Zhang, Xiaolin Yang, Dahai Yan, Xian Hou, Ming Zhou, and Guobao
Zhang for useful discussions.
MJJ acknowledges support from the Oxford Hintze Centre for Astrophysical Surveys which
is funded through generous support from the Hintze Family Charitable Foundation.
The authors gratefully acknowledge the computing time granted by the Yunnan Observatories, and provided on
the facilities at the Yunnan Observatories Super-computing Platform.

%

\vspace{5mm}
\software{BivTrunc \citep{2007ApJ...661..703S}, CosmoMC \citep{2002PhRvD..66j3511L}, emcee \citep{2013PASP..125..306F},
          fgivenx \citep{H2018}, GetDist \citep{2019arXiv191013970L}, QUADPACK \citep{1983qspa.book.....P}
          }




\clearpage

\appendix

\section{The transformation-reflection adaptive KDE method}\label{A1}
The transformation-reflection adaptive KDE is
\begin{eqnarray}
\label{kde_tra}
\hat{f}_{\mathrm{tra}}(x,y)=\frac{1}{n} \sum_{j=1}^{n}  \frac{1}{h_1(x_j,y_j)h_2(x_j,y_j)}
\left \{ K \left(\frac{x\!-\!x_j}{h_1(x_j,y_j)},\frac{y\!-\!y_j}{h_2(x_j,y_j)} \right ) \!+\! K \left (\frac{x\!-\!x_j}{h_1(x_j,y_j)},\frac{y\!+\!y_j}{h_2(x_j,y_j)} \right) \right \}.
\end{eqnarray}
where $h_1(x_j,y_j)$ and $h_2(x_j,y_j)$ are calculated via Equation (\ref{h1h2x}). The leave-one-out estimator is
\begin{eqnarray}
\label{kde_trai}
\hat{f}_{\mathrm{tra},-i}(x_i,y_i)=\frac{2}{(2n-1)}
\left \{ \! \sum_{j=1 \atop j\neq i}^{n} \!  \frac{K \left(\frac{x_i - x_j}{h_1(x_j,y_j)},\frac{y_i - y_j}{h_2(x_j,y_j)} \right)}{h_1(x_j,y_j)h_2(x_j,y_j)}
\!+\! \sum_{j=1}^{n} \!  \frac{K \left(\frac{x_i - x_j}{h_1(x_j,y_j)},\frac{y_i + y_j}{h_2(x_j,y_j)}\right)}{h_1(x_j,y_j)h_2(x_j,y_j)} \!\right\}.
\end{eqnarray}
Then by transforming back to the density of original data set, we have
\begin{eqnarray}
\label{pp_tra}
\hat{p}_{\mathrm{tra}}(z,L|h_{10},h_{20},\beta)=\frac{\hat{f}_{\mathrm{tra}}(x,y)}{(z \! + \! \tilde{\delta}_1)}, ~\mathrm{and}, ~~
\hat{p}_{\mathrm{tra},-i}(z_i,L_i|h_{10},h_{20},\beta)=\frac{\hat{f}_{\mathrm{tra},-i}(x_i,y_i)}{(z_i \!+\!\tilde{\delta}_1)}.
\end{eqnarray}
Inserting Equation (\ref{pp_tra}) to (\ref{likelihood3}), we can obtain the likelihood function for the transformation-reflection adaptive KDE estimator:
\begin{eqnarray}
\label{likelihood4}
S=-2\sum_{i}^{n}\ln[\hat{p}_{\mathrm{tra},-i}(z_i,L_i|h_{10},h_{20},\beta)] ~
+ 2n\!\!\int_{0}^{z_2}\!\!\!\!\int_{\mathrm{max}[L_1,f_{\mathrm{lim}}(z)]}^{L_2}\hat{p}_{\mathrm{tra}}(z,L|h_{10},h_{20},\beta)dzdL.
\end{eqnarray}
where $(0,z_2)$ and $(L_1,L_2)$ are redshift and luminosity limits of the simulated sample.
By numerically minimizing the object function $S$, we can determine the optimal values for $h_{10}$, $h_{20}$, and $\beta$.
Alternatively, by combining with uniform priors for $h_{10}$, $h_{20}$ and $\beta$, one can employ the MCMC algorithm to
perform Bayesian inference. Finally, we obtain the LF estimated by the transformation-reflection adaptive KDE approach,
\begin{eqnarray}
\label{phi_tra}
\hat{\phi}_{\mathrm{tra}}(z,L)=\hat{p}_{\mathrm{tra}}(z,L|h_{10},h_{20},\beta)n (\Omega\frac{dV}{dz})^{-1}.
\end{eqnarray}

\section{The KDE method considering the weighting due to the selection function}
\label{A2}

For the $i$th object in the quasar sample, with a reshift of $z_i$ and absolute magnitude of $M_i$, its weight is $w_i$.
The value of weight are given to be the inverse of the selection function for the data pair $(z_i, M_i)$.
Intuitively, a object with selection function of 0.5 is ``like'' two observations at that location \citep{2006AJ....131.2766R}.
For the values of $w_i$, we use the calculation by ``BivTrunc'', a public R wrapper of \citet{2007ApJ...661..703S}.
To perform the $\hat{\phi}_{\mathrm{tra}}$, one need to first employ the transformation-reflection method.
After transforming the original data using Equation (\ref{trans_M}), the KDE to the new data set is
\begin{eqnarray}
\label{kde_rw}
\hat{f}_{\mathrm{tr}}(x,y)=\frac{2}{2N_{\mathrm{eff}}h_1h_2}
\sum_{j=1}^{n} w_j\left( K(\frac{x\!-\!x_j}{h_1},\frac{y\!-\!y_j}{h_2}) \!+\! K(\frac{x\!-\!x_j}{h_1},\frac{y\!+\!y_j}{h_2}) \right ),
\end{eqnarray}
where $N_{\mathrm{eff}}$ is the effective sample size given by $N_{\mathrm{eff}}=\sum_{i=1}^{n} w_i$. The corresponding leave-one-out estimator is
\begin{eqnarray}
\label{kde_riw}
\hat{f}_{\mathrm{tr},-i}(x_i,y_i)=\frac{2}{(2N_{\mathrm{eff}}-w_i)h_1h_2}
\left(\! \sum_{j=1 \atop j\neq i}^{n} \! w_j K(\frac{x_i\!-\!x_j}{h_1},\frac{y_i\!-\!y_j}{h_2}) \!+\! \sum_{j=1}^{n} \! w_j K(\frac{x_i\!-\!x_j}{h_1},\frac{y_i\!+\!y_j}{h_2}) \!\right).
\end{eqnarray}
The likelihood function for the transformation-reflection KDE estimator is
\begin{eqnarray}
\label{likelihood5}
S=-2\sum_{i}^{n}\ln[\hat{p}_{\mathrm{tr},-i}(z_i,M_i|h_{1},h_{2},\delta_1)] ~
+ 2N_{\mathrm{eff}}\!\!\int_{z_1}^{z_2}\!\!\!\!\int_{M_1}^{f_{\mathrm{lim}}(z)}\hat{p}_{\mathrm{tr}}(z,M|h_{1},h_{2},\delta_1)dzdM,
\end{eqnarray}
where $z_1=0.1$, $z_2=5.3$ and $M_1=-30.7$. $\hat{p}_{\mathrm{tr},-i}$ and $\hat{p}_{\mathrm{tr}}$ are constructed similarly to Equation (\ref{pp_tr}).
By numerically minimizing the object function $S$, we can obtain the optimal bandwidths and the transformation parameter $\tilde{h}_1$, $\tilde{h}_2$, and $\tilde{\delta}_1$.
Then we follow the steps introduced in Section \ref{adaptive_kde} to achieve the adaptive KDE. Only the weight need to be involved and
the new equation is
\begin{eqnarray}
\label{kde_tra_wi}
\hat{f}_{\mathrm{tra}}(x,y)=\frac{1}{N_{\mathrm{eff}}} \sum_{j=1}^{n}  \frac{w_j}{h_1(x_j,y_j)h_2(x_j,y_j)}
\left \{ K \left(\frac{x\!-\!x_j}{h_1(x_j,y_j)},\frac{y\!-\!y_j}{h_2(x_j,y_j)} \right ) \!+\! K \left (\frac{x\!-\!x_j}{h_1(x_j,y_j)},\frac{y\!+\!y_j}{h_2(x_j,y_j)} \right) \right \}.
\end{eqnarray}
where $h_1(x_j,y_j)$ and $h_2(x_j,y_j)$ are calculated via Equation (\ref{h1h2x}). The leave-one-out estimator is
\begin{eqnarray}
\label{kde_trai_wi}
\hat{f}_{\mathrm{tra},-i}(x_i,y_i)=\frac{2}{(2N_{\mathrm{eff}}-w_i)}
\left \{ \! \sum_{j=1 \atop j\neq i}^{n} \!  \frac{w_j K \left(\frac{x_i - x_j}{h_1(x_j,y_j)},\frac{y_i - y_j}{h_2(x_j,y_j)} \right)}{h_1(x_j,y_j)h_2(x_j,y_j)}
\!+\! \sum_{j=1}^{n} \!  \frac{w_j K \left(\frac{x_i - x_j}{h_1(x_j,y_j)},\frac{y_i + y_j}{h_2(x_j,y_j)}\right)}{h_1(x_j,y_j)h_2(x_j,y_j)} \!\right\}.
\end{eqnarray}
Referring to Equations (\ref{likelihood4}) and (\ref{likelihood5}), it is easy to obtain the weighted version of likelihood function for the $\hat{\phi}_{\mathrm{tra}}$ estimator.
Finally, we obtain the quasar LF estimated by the $\hat{\phi}_{\mathrm{tra}}$ estimator,
\begin{eqnarray}
\label{phi_tra_w}
\hat{\phi}_{\mathrm{tra}}(z,M)=\hat{p}_{\mathrm{tra}}(z,L|h_{10},h_{20},\beta)N_{\mathrm{eff}} (\Omega\frac{dV}{dz})^{-1}.
\end{eqnarray}

\section{The transformation-reflection KDE for trivariate LFs}
\label{trivariate_KDE}
After transforming the original data via Equation (\ref{trans_3d}), the KDE to the density of new data ($\alpha,x,y$) is
\begin{eqnarray}
\label{kde_r_3d}
\begin{aligned}
\hat{f}_{\mathrm{tr}}(\alpha,x,y)=\frac{2}{2nh_1h_2h_3}
\sum_{j=1}^{n} \left( K(\frac{\alpha\!-\!\alpha_j}{h_1},\frac{x\!-\!x_j}{h_2},\frac{y\!-\!y_j}{h_3})
\!+\! K(\frac{\alpha\!-\!\alpha_j}{h_1},\frac{x\!-\!x_j}{h_2},\frac{y\!+\!y_j}{h_3}) \right ).
\end{aligned}
\end{eqnarray}
The corresponding leave-one-out estimator is
\begin{eqnarray}
\label{kde_ri_3d}
\begin{aligned}
\hat{f}_{\mathrm{tr},-i}(\alpha_i,x_i,y_i)=\frac{2}{(2n-1)h_1 h_2 h_3}
\left(\! \sum_{j=1 \atop j\neq i}^{n} \! K(\frac{\alpha_i\!-\!\alpha_j}{h_1}, \frac{x_i\!-\!x_j}{h_2}, \frac{y_i\!-\!y_j}{h_3})
\!+\! \sum_{j=1}^{n} \!K(\frac{\alpha_i\!-\!\alpha_j}{h_1}, \frac{x_i\!-\!x_j}{h_2}, \frac{y_i\!+\!y_j}{h_3}) \!\right).
\end{aligned}
\end{eqnarray}
The determinant of Jacobian matrix for the transformation by Equation (\ref{trans_3d}) is $\mathrm{det}(\mathbf{J})=\frac{1}{(z+\delta_1)}$.
Thus by transforming back to the density of original data set, we have
\begin{eqnarray}
\label{pp_tr_3d}
\hat{p}_{\mathrm{tr}}(\alpha,z,L|h_{1},h_{2},h_{3},\delta_1)=\frac{\hat{f}_{\mathrm{tr}}(\alpha,x,y)}{(z \! + \! \delta_1)}, ~\mathrm{and}, ~~
\hat{p}_{\mathrm{tr},-i}(\alpha_i,z_i,L_i|h_{1},h_{2},h_{3},\delta_1)=\frac{\hat{f}_{\mathrm{tra},-i}(\alpha_i,x_i,y_i)}{(z_i \!+\!\delta_1)}.
\end{eqnarray}
Inserting Equation (\ref{pp_tr_3d}) to (\ref{likelihood3}), we can obtain the likelihood function:
\begin{eqnarray}
\label{likelihood_3d}
S=-2\sum_{i}^{n}\ln[\hat{p}_{\mathrm{tr},-i}(\alpha_i,z_i,L_i|h_{1},h_{2},h_{3},\delta_1)] ~
+ 2n\!\!\int_{\alpha_1}^{\alpha_2}\!\!\!\!\int_{0}^{z_2}\!\!\!\!\int_{\mathrm{max}[L_1,f_{\mathrm{lim}}(z)]}^{L_2}\hat{p}_{\mathrm{tr}}(\alpha,z,L|h_{1},h_{2},h_{3},\delta_1)d\alpha dzdL.
\end{eqnarray}
where $(\alpha_1,\alpha_2)$, $(0,z_2)$ and $(L_1,L_2)$ are spectral index, redshift and luminosity limits of the sample.
Finally, we can obtain the trivariate LF estimated by the transformation-reflection approach,
\begin{eqnarray}
\label{phi_tr_3d}
\hat{\Phi}_{\mathrm{tr}}(\alpha,z,L)=\hat{p}_{\mathrm{tr}}(\alpha,z,L|h_1,h_2,h_3,\delta_1)n (\Omega\frac{dV}{dz})^{-1}.
\end{eqnarray}
Usually, we are more interested in the bivariate LF and it can be obtained by
\begin{eqnarray}
\label{phi_tr_2d}
\hat{\phi}_{\mathrm{tr}}(z,L)=\int_{\alpha_1}^{\alpha_2} \hat{\Phi}_{\mathrm{tr}}(\alpha,z,L)d\alpha.
\end{eqnarray}





\end{document}